\documentclass[sigplan,screen]{acmart}
\usepackage{booktabs} 
\usepackage{url}
\usepackage{color}
\usepackage{colortbl}
\usepackage{xspace}
\usepackage{booktabs}
\usepackage[american]{babel}
\usepackage{listings}
\usepackage{flushend}
\usepackage{multirow}
\usepackage{amsmath}
\usepackage{setspace}
\usepackage{algorithm}
\usepackage{algorithmic}
\usepackage{listings}
\usepackage{subfigure}
\usepackage{ragged2e}
\usepackage{amsmath,bm}
\usepackage{amsmath,amssymb,amsfonts}
\usepackage{algorithmic}
\usepackage{graphicx}
\usepackage{textcomp}
\usepackage{tabularx}
\usepackage{babel}
\usepackage{tablefootnote}
\usepackage{booktabs}
\usepackage{enumitem}
\usepackage{balance}

\AtBeginDocument{%
  \providecommand\BibTeX{{%
    \normalfont B\kern-0.5em{\scshape i\kern-0.25em b}\kern-0.8em\TeX}}}

\newcommand\cparagraph[1]{\vspace{1.2mm}\noindent\textbf{#1.}}

\newcommand\SystemName{\textsc{Prom}\xspace}

\newcommand\bparagraph[1]{\vspace{1mm}\noindent\textbf{\textit{#1.}}}
\definecolor{Gray}{gray}{0.95}

\lstdefinestyle{interfaces}{
  float=tp,
  floatplacement=tbp,
  abovecaptionskip=-5pt
}

\lstdefinelanguage
   [x64]{Assembler}     
   {morekeywords={CDQE,CQO,CMPSQ,CMPXCHG16B,JRCXZ,LODSQ,MOVSXD, %
                  POPFQ,PUSHFQ,SCASQ,STOSQ,IRETQ,RDTSCP,SWAPGS, %
                  _Z3p23i,callq,
                  r8,r8d,r8w,r8b,r9,r9d,r9w,r9b, %
                  r10,r10d,r10w,r10b,r11,r11d,r11w,r11b, %
                  r12,r12d,r12w,r12b,r13,r13d,r13w,r13b, %
                  r14,r14d,r14w,r14b,r15,r15d,r15w,r15b, popcntl}}

\newcommand\Vuldeepecker{\textsc{Vulde}\xspace}
\newcommand\Mapie{\textsc{Mapie}\xspace}
\newcommand\Puncc{\textsc{Puncc}\xspace}

\newcommand\DeepTune{\textsc{DeepTune}\xspace}
\newcommand\KStock{\textsc{K.Stock \emph{et al.}}\xspace}
\newcommand\etal{\emph{et al.}\xspace}
\newcommand\CodeXGLUE{\textsc{CodeXGLUE}\xspace}
\newcommand\LineVul{\textsc{LineVul}\xspace}

\newcommand\Programl{\textsc{Programl}\xspace}
\newcommand\tlp{\textsc{Tlp}\xspace}

\newcommand\RAISE{\textsc{RISE}\xspace}
\newcommand\TESSERACT{\textsc{TESSERACT}\xspace}

\newcommand\IR{\textsc{IR2Vec}\xspace}

\setcopyright{cc}
\setcctype{by}
\acmDOI{10.1145/3696443.3708959}
\acmYear{2025}
\copyrightyear{2025}
\acmISBN{979-8-4007-1275-3/25/03}
\acmConference[CGO '25]{Proceedings of the 23rd ACM/IEEE International Symposium on Code Generation and Optimization}{March 01--05, 2025}{Las Vegas, NV, USA}
\acmBooktitle{Proceedings of the 23rd ACM/IEEE International Symposium on Code Generation and Optimization (CGO '25), March 01--05, 2025, Las Vegas, NV, USA}
\received{2024-09-12}
\received[accepted]{2024-11-04}

\begin{document}

\title{Enhancing Deployment-Time Predictive Model Robustness for Code Analysis and Optimization}

\author{Huanting Wang}
\orcid{0000-0003-0579-4295}
\affiliation{%
  \institution{School of Computer Science, University of Leeds}
  \city{Leeds}
  \country{United Kingdom}
}
\email{schwa@leeds.ac.uk}

\author{Patrick Lenihan}
\orcid{0009-0005-6816-353X}
\affiliation{%
  \institution{School of Computer Science, University of Leeds}
  \city{Leeds}
  \country{United Kingdom}
}
\email{p.j.lenihan1@leeds.ac.uk}

\author{Zheng Wang}
\orcid{0000-0001-6157-0662}
\affiliation{%
  \institution{School of Computer Science, University of Leeds}
  \city{Leeds}
  \country{United Kingdom}
}
\email{z.wang5@leeds.ac.uk}

\begin{abstract}
Supervised machine learning techniques have shown promising results in code analysis and optimization problems. However, a learning-based solution can be brittle because minor changes in hardware or application workloads -- such as facing a new CPU architecture or code pattern -- may jeopardize decision accuracy, ultimately undermining model robustness. We introduce \SystemName, an open-source library to enhance the robustness and performance of predictive models against such changes during \emph{deployment}. \SystemName achieves this by using statistical assessments to identify test samples prone to mispredictions and using feedback on these samples to improve a deployed model.
We showcase \SystemName by applying it to 13 representative machine learning models across 5 code analysis and optimization tasks.  Our extensive evaluation demonstrates that \SystemName can successfully identify an average of 96\% (up to 100\%) of mispredictions. By relabeling up to 5\% of the \SystemName-identified samples through incremental learning, \SystemName can help a deployed model achieve a performance comparable to that attained during its model training phase. 
\end{abstract}



    \maketitle

\section{INTRODUCTION}

Supervised machine-learning (\texttt{ML}) is a powerful tool for code analysis and optimization tasks like bug detection~\cite{Wang2020FUNDED,fu2022linevul,livuldeepecker,wang2024combining} and
runtime- or compiler-based code optimization~\cite{wang2009mapping,tournavitis2009towards,wang2010partitioning,cummins2017end,POEM,cummins2021programl,zhai2023tlp,ren2017optimise,wang2024drlcap,wang2022automating,ren2021adaptive}. ML works by training a predictive model from training samples and then applying the trained model to \emph{previously unseen} programs within operational environments that often have diverse application workloads and hardware~\cite{grewe2013portable,wang2018machine}. 

ML solutions, while powerful, can be fragile. Small changes in hardware or application workloads can reduce decision accuracy and model robustness~\cite{sculley2014machine}. This often arises from ``\emph{data drift}''~\cite{quinonero2008dataset,tsymbal2004problem}, where the training and test data distributions no longer align. Data drift can occur when the assumption that past training data reflects future test data is violated. In code optimization, this can result from changes in workload patterns, runtime libraries, or hardware micro-architectures. It is a particular challenge for ML-based performance optimizations, where obtaining sufficient performance training data is difficult~\cite{ogilvie2017minimizing, cummins2017synthesizing}.


Existing efforts to enhance ML robustness for code optimization have predominantly focused on improving the learning efficiency or model generalization during \emph{the design time}. These approaches include synthesizing benchmarks to increase the training data size \cite{cummins2017synthesizing, catak2021data, tsimpourlas2023benchdirect}, finding better program representations~\cite{POEM,cummins2021programl,Wang2020FUNDED,wang2024combining}, and combining multiple models to increase the model's generalization ability~\cite{mc,taylor2018adaptive,Wang2020FUNDED}. While important, these design-time methods are unlikely to account for all potential changes during deployment~\cite{mallick2022matchmaker}. Although there has been limited exploration into the validation of model assumptions~\cite{EmaniO15} for runtime scheduling, existing solutions assume a specific ML architecture and lack generalizability.


We introduce \SystemName, an open-source toolkit designed to address data drift \textit{during deployment}, specifically targeting code optimization and analysis tasks. \SystemName is not intended to replace design-time solutions but to offer a complementary approach to improve ML robustness during deployment. Its primary objective is to ensure the reliability of an already deployed system in the face of changes and support continuous improvements in the end-user environment.
To this end, \SystemName offers a Python interface for training and deploying supervised ML models, focusing on detecting data drift post-deployment. Model and application developers can integrate \SystemName into the ML workflow by implementing its abstract class, usually requiring just a few dozen lines of code.

\SystemName adopts the emerging paradigm of \emph{prediction with rejections}~\cite{sec2017Transcend, tang2014highly, barbero2020transcending, RISE}, which identifies instances where predictions may be inaccurate, allowing for corrective measures when data drift occurs.
For instance,  an ML-based performance tuner can notify users when the model prediction, like the compiler flags to be used for a given program, might not yield good performance, prompting them to use alternative search processes~\cite{ansel:pact:2014,wang2022automating} to find better solutions~\cite{haneda2006impact}. Likewise, a bug detector can alert users to potential false positives for expert inspection.
Essentially, this capability allows using alternative metrics when predictive model performance deteriorates. By flagging likely mispredictions, \SystemName supports continuous learning by using these mispredicted instances as additional training samples to enhance model performance in a production environment.

To evaluate whether a predictive model may mispredict a test input, \SystemName computes the \emph{credibility} and \emph{confidence} scores of the prediction during deployment. Credibility measures the likelihood that a prediction aligns with the learned patterns. High credibility means the test input is highly consistent with the training data, suggesting the model's prediction is likely to be reliable. Conversely, the confidence score estimates the model's certainty in its prediction. \SystemName uses the two scores to determine whether the model's outcome should be accepted or requires further investigation. Our intuition is that a prediction is reliable only if the model shows high confidence in its predictions and these predictions, along with the model's confidence level, are credible.

\SystemName employs conformal prediction (\texttt{CP}) theory \cite{balasubramanian2014conformal}  to assess the test input's \emph{nonconformity} to compute the confidence and credibility scores of a prediction. Nonconformity is measured by comparing the test input against samples from a \emph{calibration dataset} held out from the model training samples. The idea is to observe the ML model's performance on the calibration set and then evaluate the test input's similarity (or ``strangeness") to the calibration samples. 

\SystemName draws inspiration from recent advances in applying CP to detect drifting samples in malware prediction~\cite{barbero2020transcending, sec2017Transcend} and wireless sensing~\cite{RISE}. While CP has shown promise in these domains, its effectiveness for code optimization tasks remains unclear. This paper presents the first application of CP to code optimization tasks like loop vectorization and neural network code generation. Doing so requires addressing several key limitations of existing approaches. One major drawback of prior methods is they rely on the entire calibration dataset to compute the nonconformity of test samples, which is ill-suited for code with diverse patterns. For example, if we want to estimate the model's accuracy on a computation-intensive program, including many calibration samples with different characteristics (e.g., memory-bound benchmarks) can mislead and bias the credibility estimation. Furthermore, previous solutions employ a monolithic nonconformity function that lacks robustness across different ML models and tasks. They also do not support regression methods and usually require changing the underlying ML model~\cite{RISE}. \SystemName is designed to overcome these pitfalls.


Unlike prior work~\cite{barbero2020transcending, sec2017Transcend, RISE}, \SystemName adopts an adaptive scheme to measure a test sample's nonconformity. Instead of using the entire calibration dataset, \SystemName dynamically selects a subset of calibration samples with similar characteristics to the test sample in the feature space defined by the ML model. 
When computing the nonconformity score, \SystemName assigns different weights to these selected samples based on their distance from the test sample, giving higher weight to closer samples. This scheme allows \SystemName to construct a calibration set that closely matches the test sample distribution, thereby improving nonconformity estimation accuracy. 

\SystemName improves conformity reliability by using multiple statistical functions to compute nonconformity scores and applying a majority voting scheme to approve or reject predictions. It is extensible, allowing easy addition of new nonconformity functions. \SystemName also supports regression by combining CP with clustering algorithms. Unlike \cite{RISE}, it uses a model-free approach instead of learning a probabilistic classifier for data drift detection.



As a potential mitigation of data drift, we showcase that \SystemName enables a learning-based method to enhance its robustness and maintain reliable performance over time. This is achieved by adopting incremental learning \cite{gepperth2016incremental, ade2013methods} to retrain a deployed model using drifting samples from the production environment. Depending on the specific application, one can relabel a few test samples flagged by \SystemName and use the relabeled data to retrain the model, by only seeking feedback and user intervention on instances showing data drift. 
We stress that incremental learning is just one of the possible remedies, but such mitigations hinge on accurately detecting drifted samples - the main focus of this work.

We demonstrate the utility of \SystemName by applying it to 13 representative ML methods developed by independent researchers for code optimization and analysis~\cite{magni2014automatic,cummins2017end,brauckmann2020compiler,cummins2021programl,POEM,KStock,haj2020neurovectorizer,
livuldeepecker,lu2021codexglue,fu2022linevul,Wang2020FUNDED,zhai2023tlp}. Our case studies cover five problems, including heterogeneous device mapping~\cite{brauckmann2020compiler,cummins2021programl,cummins2017end,POEM}, GPU thread coarsening~\cite{magni2014automatic,cummins2017end,POEM}, loop vectorization~\cite{POEM,KStock}, neural network code generation~\cite{zhai2023tlp} and source-code level bug detection~\cite{Wang2020FUNDED,lu2021codexglue,fu2022linevul,livuldeepecker}.
Experimental results show that \SystemName can successfully identify an average of $96\%$ (up to 100\%) of test inputs where the underlying ML model mispredicts with a false-positive rate\footnote{This occurs when the ML model gives an accurate prediction, but \SystemName believes otherwise.} of less than 14\%. By employing incremental learning, \SystemName substantially enhances prediction performance in the operational environment, allowing the deployed model to match the performance achieved during its design phase. Notably, this improvement is achieved with minimal user intervention or profiling overhead. In our evaluation, \SystemName requires relabeling fewer than $5\%$ of the samples identified as drifted by \SystemName, which are then used to update the model through retraining.

This paper makes the following contributions:
\begin{itemize}

\item The first framework to address ML model reliability after model deployment for code analysis and optimization;

\item A collection of ready-to-use conformity measurements to support \emph{both} classification and regression models for code optimization and analysis during deployment (Sec.~\ref{section:overview});
\item A new adaptive weighting scheme and ensemble approach when applying CP to detect data drift (Sec.~\ref{sec:cpv}) and an ensemble approach to detect mispredictions;
\item A large-scale study validating the effectiveness of CP in code optimization and analysis tasks (Sec.~\ref{section:results}).
 \end{itemize}

\begin{figure}[t!]
    \subfigure[Data drift leads to deteriorating performance in software vulnerability detection.]{
    \begin{minipage}[t]{0.9\columnwidth}
    \centering
    \includegraphics[width=\columnwidth]{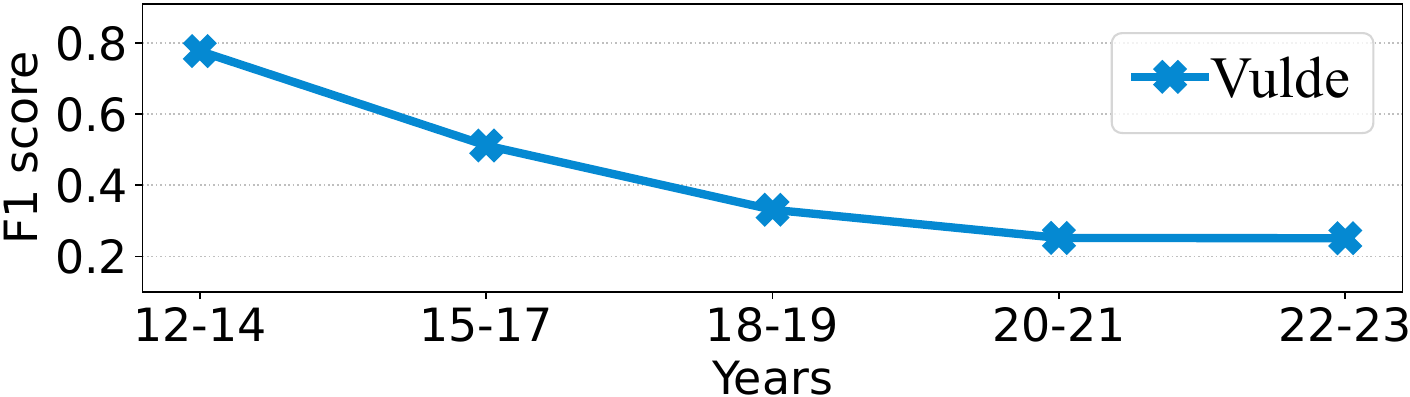}
    \end{minipage}
    \label{fig:motivation}
    }
        \quad
    \subfigure[CVE-2012-4559: A ``\emph{double-free}" for \texttt{name} at lines 6 and 8.]{

    \begin{minipage}[t]{0.95\columnwidth}
    \centering
\begin{lstlisting}[language=C, basicstyle=\scriptsize\ttfamily]
static sftp_parse_attr_3(...) {
ssh_string name = NULL;
...
if ((name = buffer_get_ssh_string(buf))...) 
    {...}
ssh_string_free(name);
...
ssh_string_free(name);}
\end{lstlisting}
    \end{minipage}
    \label{fig:motivation_lst1}
    }
     \quad
    \subfigure[CVE-2023-27537: A potential ``\emph{double-free}" due to multiple concurrent threads can invoke \texttt{hsts\_free{}} at line 6 at the same time.]{

    \begin{minipage}[t]{0.95\columnwidth}
    \centering
\begin{lstlisting}[language=C, basicstyle=\scriptsize\ttfamily]
#define NUMT 100

CURLcode curl_easy_cleanup(...){
sts = ...;
if(sts) {...
   hsts_free(sts);}}
...
static void* pull_one_url(...){
for (i = 0; i < NUMT; i++) {
    CURL* curl = ...;
    ...
    curl_easy_cleanup(curl); }}
   
int main(...){...
for (i = 0; i < NUMT; i++) {
    pthread_create(pull_one_url,...);
...}
\end{lstlisting}
    \end{minipage}
    \label{fig:motivation_lst2}
    }
    
    \centering
    \caption{Motivation example: impact of data drift on ML models for code vulnerability detection.}
    \label{fig:roc}
\end{figure}

\section{MOTIVATION}~\label{sec:motivation}

As a motivating example, consider training and using an ML model to detect source code bugs. In this pilot study, we consider  \Vuldeepecker~\cite{livuldeepecker}, which uses a long short-term memory (LSTM) network for bug detection. Following the original setup, we train and test the model on labeled samples from the common vulnerabilities and exposures (CVE) dataset. We use the open-source release of \Vuldeepecker and ensure results are comparable to those in the source publication of \Vuldeepecker.


Figure~\ref{fig:motivation} shows what happens if we train \Vuldeepecker using CVE data collected between 2012 and 2014 and then apply it to real-life code samples developed between 2015 and 2023. This mimics a scenario where the code and bug patterns may evolve after a trained model is deployed.
The trained model achieves an F1 score of more than 0.8 (ranging from 0 to 1, with higher being better) when the test and training data are collected from the same time window. However, the F1 score drops to less than 0.3 when tested on samples collected from future time windows. This shows that data drift can severely impact model performance, which was also reported in prior studies~\cite{sec2019TESSERACT, Wang2020FUNDED}.

To illustrate code pattern changes, Figures~\ref{fig:motivation_lst1} and \ref{fig:motivation_lst2} show two cases of ``\emph{double-free}'' vulnerabilities, one from 2012 and one from 2023. Earlier cases were simpler, where the same memory was freed twice (e.g., \texttt{name} is freed on lines 6 and 8). A model trained on these samples is unlikely to detect the later, more complex case in Figure~\ref{fig:motivation_lst2}, where a ``double-free'' occurs due to concurrent threads calling the same buffer-free routine. Because it is difficult to collect a training dataset which covers all possible code patterns seen in deployment time, data drift can happen in the real-life deployment of ML models for code-related tasks.

\section{BACKGROUND}
\subsection{The Need of Credibility Evaluation\label{sec:cen}}
To detect unreliable prediction outcomes, a straightforward approach is to analyze probability distribution given by an ML model. For instance, a high prediction probability for a specific label might suggest high confidence in the prediction. However, this alone does not fully capture the prediction's reliability, as ML models can produce skewed probabilities for rarely seen samples. Consider multi-class classification for example. An ML model predicts the likelihood, $r^i$, that a given input belongs to each class ($c_1$ to $c_n$). However, if the input's pattern significantly differs from training samples, the model might assign a low probability to some classes (e.g., $r^1 \approx 0.0$ for class $c_1$). This can disproportionately inflate the probabilities of other classes ($r^2$ to $r^n$) as the sum of probabilities needs to equal 1.0~\cite{bishop2006pattern}.  In this case, a high probability does not equate to high prediction confidence. Therefore, assessing the model's credibility requires an approach that evaluates how well the input aligns with the training data.

\subsection{Statistical Assessment}

\SystemName uses statistical assessments to evaluate prediction credibility and confidence. Unlike typical probabilistic evaluations in ML models, which assess the likelihood of a test sample belonging to a certain class or value in isolation, statistical assessments draw from historical data distributions.  They answer questions like: ``\emph{How likely is the test sample to belong to a class compared to all other possible classes}''? By framing the sample within the broader context of historical decisions and probabilistic distributions, statistical assessments quantify the uncertainty of a prediction.
\subsection{Conformal Prediction\label{sec:pvalue}}

\SystemName is built on conformal prediction (CP)~\cite{balasubramanian2014conformal,angelopoulos2021gentle}, which, given a model $g$ and a significance level, defines a prediction region that contains the true value with a certain probability. CP constructs this region based on training data distribution, accounting for noise and variability. CP was designed to improve prediction coverage by calculating a prediction range. \SystemName utilizes CP, \emph{for a different purpose:} evaluating the reliability of a model prediction.

                      
\cparagraph{P-value}
\SystemName uses the p-value \cite{thisted1998p}, given by CP, to assess the prediction credibility in its predictions and the confidence of the credibility. 
We use it to quantify evidence that contradicts a null hypothesis. For instance, in determining the \emph{confidence} of a classifier's prediction, the null hypothesis would assert that there is no substantial difference between the predicted class and any other class, as observed from the model's probability distribution. Likewise, in assessing the \emph{credibility} of a prediction, the null hypothesis assumes that the prediction does not fall within a specific prediction region computed by CP.
In \SystemName (see also Sec.~\ref{sec:cpv}), a high p-value suggests that the likelihood of the observed data under the null hypothesis is small, providing strong evidence against it. Conversely, a low p-value indicates a high likelihood of the observed data under the null hypothesis, thus offering weaker evidence against it. In other words, a high p-value suggests that the prediction is reliable.

\begin{figure}[t!]
 \centering
\includegraphics[width=0.45\textwidth]{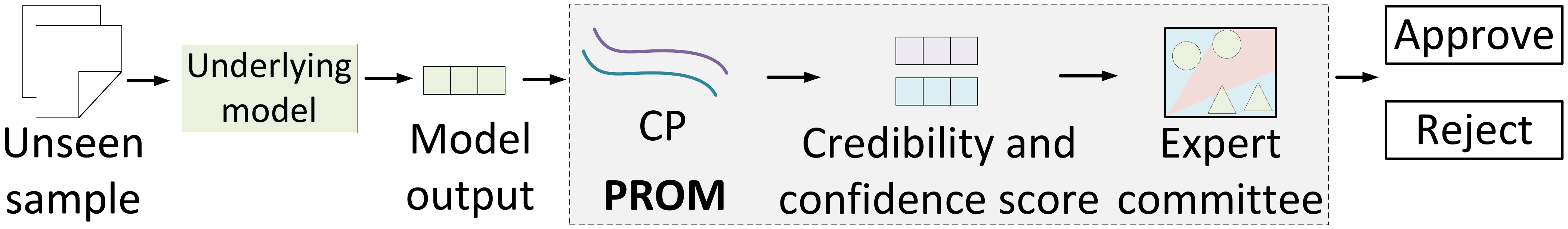}\\
  \caption{Workflow of \SystemName during deployment.}

 \label{fig:Workflow}
\end{figure}

\begin{figure}[t!]
 \centering
\includegraphics[width=0.48\textwidth]{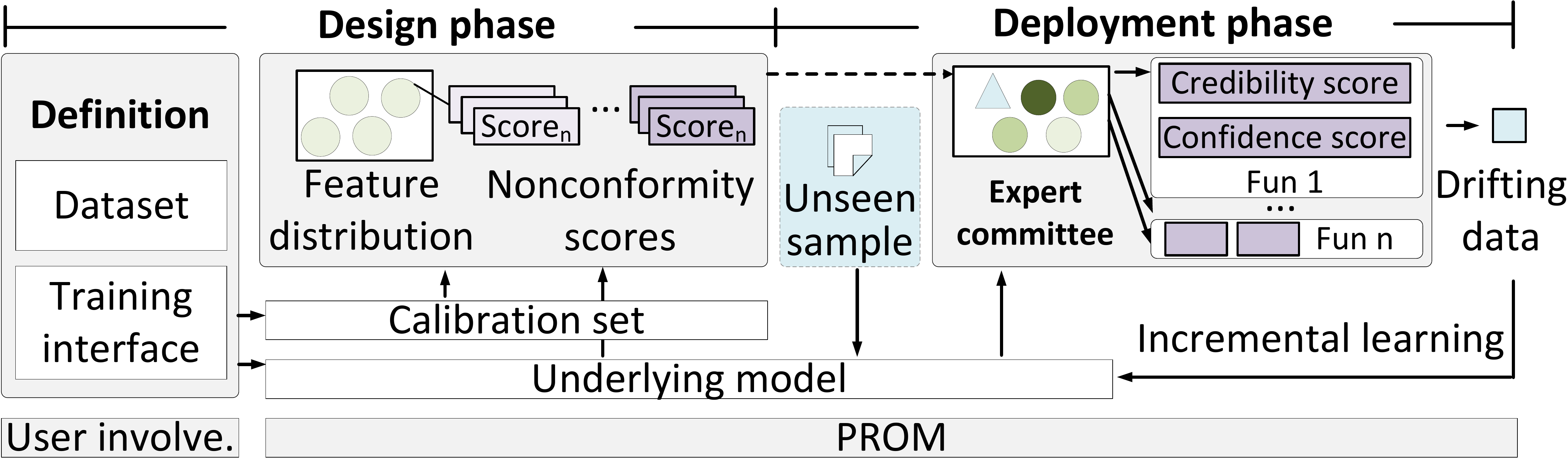}\\
  \caption{At design time, \SystemName splits the training data into training and calibration sets. During deployment, it calculates credibility and confidence scores, using majority voting to detect drifting samples. These samples can then be labeled for model updates via offline incremental training.
  }
 \label{fig:framework}
\end{figure}

\begin{figure}[t!]
\centering
\begin{lstlisting}
from prom import ModelInterface

class ModelDefinition(ModelInterface):
    def __init__(self, model, cali_dataset):
        # Setting the underlying model
        self.model = model
        self.calibration_data = cali_dataset
        super().__init__()
    
    def data_partitioning(self, dataset, calibration_ratio=0.1):
        # Splitting the training data into training and calibration sets
        ... 
        return training_data, calibration_data
        
    def predict(self, X, significant_level=0.1):
        ...
        # Underlying model prediction function also returns a probabilistic vector
        pred, probability = self.model.predict(X)
        #Call the expert committee
        drifted = self.classifier(probability, significant_level)
        return pred, drifted
    
    def feature_extraction(self, X):
        # Convert the model input into a feature vector
        ...
        return self.model.feature_extraction(X)
        
if __name__ == '__main__':
  model = ModelDefinition(mymodel, dataset)
  pred, drifted = model.predict(sys.argv[1])
\end{lstlisting}
 \caption{Simplified code template of \SystemName.}
 \label{fig:user}
\end{figure}

\section{OVERVIEW of PROM\label{section:overview}}
Figure~\ref{fig:Workflow} illustrates how \SystemName can enhance learning-based methods \emph{during deployment}. Users of \SystemName are ML model or application developers.
\SystemName requires no change to a user model's structure and working mechanism. In this paper, we refer to the user model as the ``\emph{underlying model}". 

\cparagraph{User involvement}
For a given input, the underlying model works as it would without \SystemName during inference. Users of \SystemName need only provide the model training dataset and the training interface to \SystemName.

\cparagraph{Added values of \SystemName}
\SystemName serves as an open-source framework that provides:
an adaptive scheme for constructing the calibration set; a collection of conformal prediction methods that take as input the intermediate results (e.g., probabilities of each class label produced by the underlying model) to compute a credibility and confidence score, which is used to suggest whether to accept the prediction outcome; and an ensemble approach to detect mispredictions.

\cparagraph{Scope} \SystemName goes beyond a standard CP library~\cite{mendil2023puncc,taquet2022mapie} and is not limited to code-related tasks. It tackles a key challenge in ML for code analysis and optimization - insufficient data for training robust models at design time - by offering an orthogonal approach to enhance model reliability during deployment. This makes it particularly useful for ML in code analysis and optimization.


\subsection{Implementation}
We implemented \SystemName as a Python package, offering an API for use during both the ML model design and deployment phases. 
\SystemName provides interfaces to assess framework setup (Sec.~\ref{sec:dta}), automatically searches for hyperparameter settings on the training and calibration datasets, and provides examples to showcase its utilities. These include all the case studies and ML models used in this work (Sec.~\ref{sec:setup}) and simpler examples for beginners.
\SystemName supports classification and regression methods built upon classical ML methods (e.g., support vector machines) and more recent deep neural networks. Figures \ref{fig:framework} and \ref{fig:user} provide an overview of \SystemName's role during the model design and deployment phases, described as follows.

\subsubsection{Model design phase\label{sec:mdp}}

As depicted in Figure~\ref{fig:user}, using \SystemName requires overwriting a handful of methods in \SystemName's \texttt{ModelDefinition} class and exposing the underlying model's internal outputs.

\cparagraph{Training data partitioning}
\SystemName is based on split CP~\cite{balasubramanian2014conformal,angelopoulos2021gentle}, which divides the training data into a ``\emph{training dataset}'' and a ``\emph{calibration dataset}''. 
\SystemName uses the calibration dataset to detect drifting test samples during deployment. By default, it randomly sets aside 10\% of the training data (up to 1,000 samples) for calibration, a method shown to be effective in prior work~\cite{angelopoulos2021gentle,vovk2012conditional}. \SystemName also offers a way to assess the suitability of the calibration dataset (Sec.~\ref{sec:dta}), or users can provide their own holdout calibration dataset.


\cparagraph{Changes to user models}
For classification tasks, the user model should implement a prediction function (line 15) that returns both a prediction and a probability vector. Most ML classifiers already associate probabilities with each class, which can be easily accessed. Popular frameworks like scikit-learn, PyTorch, and TensorFlow directly provide probability distributions. For example, scikit-learn's \texttt{predict\_proba} method offers probabilistic values for 33 common ML models. In neural networks, probabilities can usually be extracted from the hidden layer before the output. For regression models~\cite{angelopoulos2021gentle, romano2019conformalized}, \SystemName applies a similar approach.


\cparagraph{Feature extraction} The user needs to provide a feature extraction function to convert the model input into a feature vector of numerical values. For example, this could be a neural network to generate embeddings of the input~\cite{devlin2018bert,feng-etal-2020-codebert} or a function to summarize the input programs into numerical values like the number of instructions \cite{wang2022automating}. Since most ML models already require this function, this requirement should not incur additional engineering effort.

\cparagraph{Process calibration dataset}
The user model is trained \emph{outside} the \SystemName framework using any method the user deems appropriate. The trained model is loaded and passed as a Python object to \SystemName.
With the calibration dataset and the trained user model, \SystemName \emph{automatically}  preprocesses the calibration dataset offline before deploying the ML model (Sec.~\ref{sec:cp}). This is done by applying the learned model to each calibration sample and using a nonconformity function described in Sec.~\ref{sec:ncm} to calculate a score. This score reflects how `strange' or `\emph{non-conforming}' each calibration example is compared to the learned model. 

\cparagraph{Significant level}
The user can set a \emph{significant level}, $1-\epsilon$, to determine the severity of data drifts.
A smaller $\epsilon$ can reduce the probability of misprediction by \SystemName but can lead to a higher false positive rate. 
By default, \SystemName sets $\epsilon$ to 0.1. 

\cparagraph{Overwrite the prediction function}
As a final step, the model developer needs to overwrite a prediction function (\texttt{predict}) function to return the model's prediction for a test input. The original prediction function can be invoked as a subroutine, and the \SystemName prediction function is used during deployment.


\subsubsection{Model deployment phase\label{sec:mdeploymentp}}
During deployment, the user model functions as usual, taking a test sample and making a prediction. The difference with \SystemName is that it also suggests whether to accept or reject the prediction. The user can use this outcome to identify mispredictions and provide ground truth, and \SystemName will use these relabeled drifting samples to update the model.


\section{METHODOLOGY\label{sec:cp}}
\begin{figure}[t!]
 \centering
\includegraphics[width=0.47\textwidth]{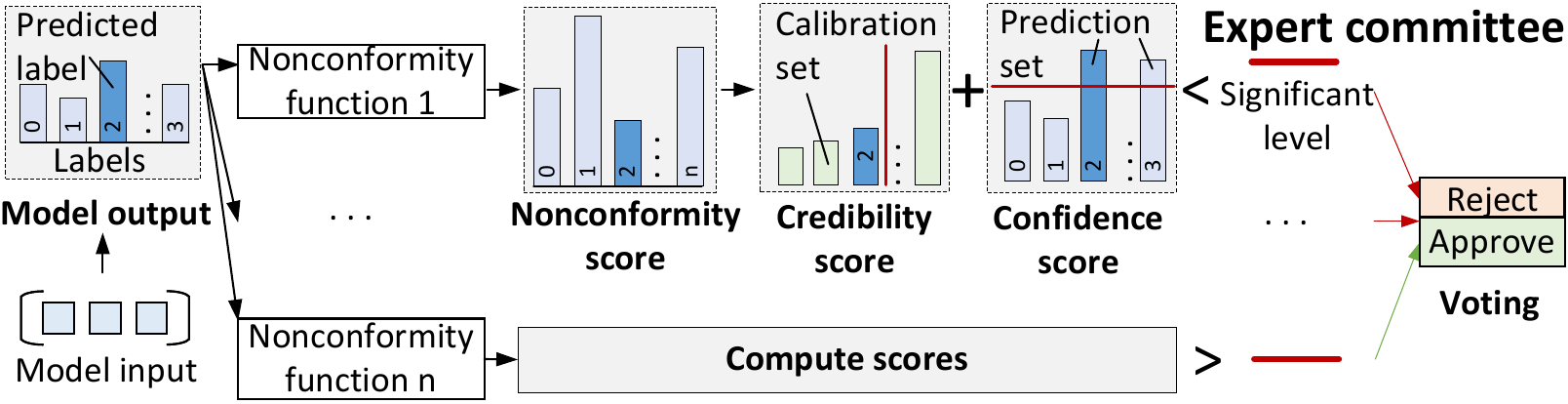}\\
  \caption{\SystemName integrates multiple nonconformity functions that vote to reject or approve the ML prediction.}
 \label{fig:Expert}
\end{figure}

As shown in Figure~\ref{fig:Expert}, during deployment, \SystemName uses multiple (default: 4) nonconformity functions to independently compute the prediction's credibility and confidence scores. These scores are compared to a pre-defined significance level, $1-\varepsilon$ (Sec.~\ref{sec:mdp}), to decide whether to accept the prediction. If both scores fall below the threshold, the test sample is flagged as drifting. The results are then aggregated using majority voting, where each nonconformity function (expert) decides whether the prediction should be accepted, forming an ensemble ``expert committee''.



\subsection{Nonconformity Measures\label{sec:nm}}
\begin{figure}[t!]
 \centering
\includegraphics[width=0.45\textwidth]{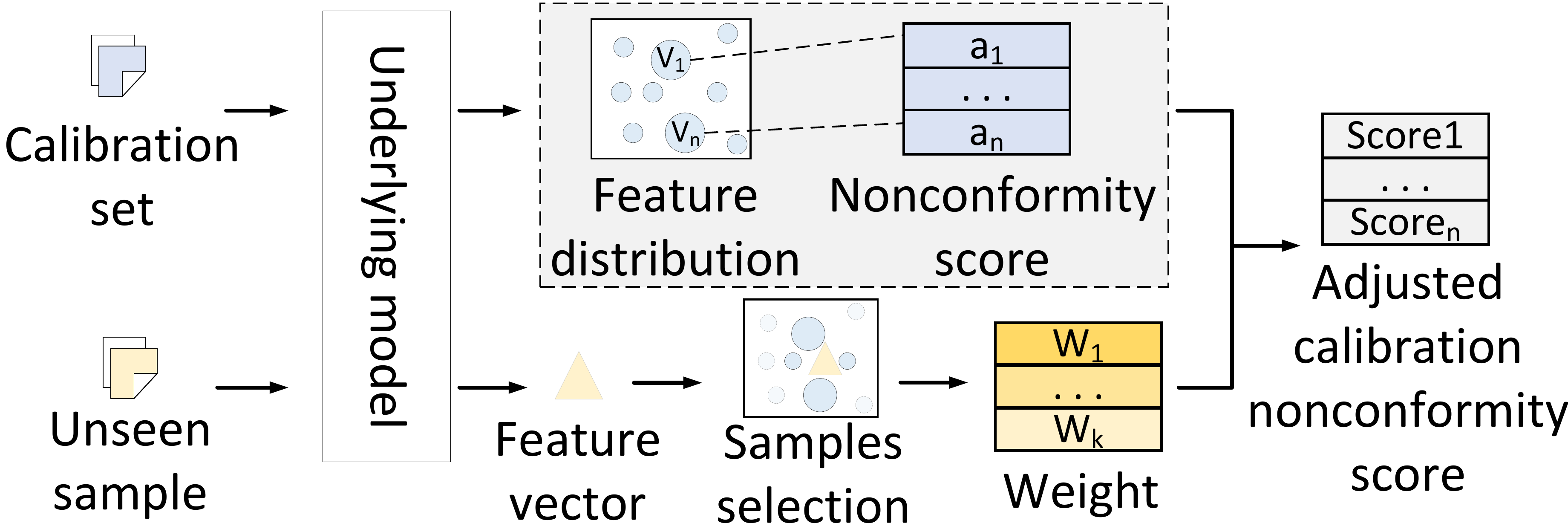}\\
  \caption{
  \SystemName dynamically selects a subset of the holdout calibration dataset to assess the test input's nonconformity. 
  }
 \label{fig:knn_ncm}
\end{figure}

\SystemName computes the p-value of a prediction using nonconformity functions, which are then used to derive credibility and confidence scores. 
Previous work on using CP to detect drifting samples~\cite{barbero2020transcending, sec2017Transcend,RISE} focuses primarily on classification tasks and does not extend to regression. \SystemName is the first framework to support both classification and regression for data drift detection. Additionally, prior methods only consider the predicted label, ignoring the probability distribution across labels, whereas \SystemName accounts for probabilities across all labels in classification tasks (Sec.~\ref{sec:cen}).



\subsubsection{Nonconformity functions\label{sec:ncm}}
\SystemName integrates multiple ready-to-use nonconformity functions, and the choice of nonconformity functions can be customized by passing a list to the relevant \SystemName interface. By default, \SystemName uses 4 nonconformity functions: \textit{LAC}~\cite{sadinle2019least}, \textit{TopK}~\cite{angelopoulos2020uncertainty}, \textit{APS}~\cite{romano2020classification} and \textit{RAPS}~\cite{angelopoulos2020uncertainty}.  Other nonconformity functions can be easily incorporated into \SystemName by implementing an abstract class.


For regression tasks, our nonconformity functions compute the nonconformity score using the residual error between the prediction and the ground truth. Since we do not have the ground truth during deployment, we approximate it using the k-nearest neighbour algorithm~\cite{ishimtsev2017conformal, cherubin2021exact}. This approximation is based on the null hypothesis that the test sample is similar to those encountered during design time.

Specifically, \SystemName finds the k-nearest neighbors (we set $k$ to be 3 in this work), denoted as $N_k({n+1})$, of $s_{n+1}$. The distance is measured by computing the Euclidean distance~\cite{danielsson1980euclidean} between the test sample $s_{n+1}$ and calibration samples on the feature space.  We then approximate the true value of $s_{n+1}$ by averaging the distance of k-nearest neighbors, $
y_{s_{n+1}} = \frac{1}{k} \sum_{i \in N_k({n+1})} y_{s_i}$. The estimated value is then passed to a regression-based nonconformity function to compute the nonconformity score of the test sample. Essentially, we approximate the ground truth by assuming the samples seen at the design time are sufficient to generate an accurate prediction. If this assumption is violated due to drifting test samples, it will likely result in a large residual error (and a greater nonconformity score).

\subsubsection{Computing p-value\label{sec:cpv}}
\SystemName uses a \emph{p-value} to assess whether a test sample $s$ fits within the prediction region defined by the calibration dataset, which reflects the training data distribution. To compute the p-value, a subset of calibration samples is selected, their nonconformity scores are adjusted, and these scores are used to derive the p-value for the model's prediction.

\cparagraph{Calibration nonconformity scores}
As shown in Figure~\ref{fig:knn_ncm}, \SystemName dynamically selects a subset of calibration samples to adjust the nonconformity score. Specifically, it computes the Euclidean distance between each calibration sample and the test input based on their feature vectors, sorting the samples by distance. By default, the closest 50\% of calibration samples are selected. If the dataset contains fewer than 200 samples, all of them are selected; a threshold that can be configured via the \SystemName API. This nearest subset of calibration samples is chosen to estimate the nonconformity of the test data relative to the training data. These distances are also used as weights to adjust nonconformity scores.

For a calibration dataset with $n$ samples, $(a_1, a_2, \ldots, a_n)$, \SystemName computes nonconformity scores and feature vectors ($v_1, v_2, \ldots, v_n$) offline. For a new test sample $s_{n+1}$, it extracts the feature vector $v_{n+1}$, calculates the distances to calibration samples, and selects $K$ samples. The weight $w_i$ for each selected sample $i$ is given by:
\begin{equation}
w_i = \exp \left( -\frac{\|v_i - v_{n+1}\|^2}{\tau} \right), i\in\{1, ..., k\}
\end{equation}
where $\|v_i - v_{n+1}\|^2$ is the l2-norm, and $\tau$ is a temperature hyperparameter (default 500). This weight is used to adjust the nonconformity score: $a_i = w_i \times a_i$.

\cparagraph{P-value for classification}
After selecting the calibration samples and adjusting their nonconformity scores, \SystemName calculates the p-value for each test sample. First, it determines the nonconformity score $a_{n+1}^{y^p}$ for the predicted outcome $y^p$. Then, it evaluates the similarity of the test sample to the chosen calibration samples to compute the p-value, $p_{s_i}$, as:
\begin{equation}
\footnotesize
p_{s_i} =\frac{\text {COUNT}\left\{i \in\{1, ..., n\}: y_{i}=y^{p} \text { and } a_{i}^{y^{p}} \geq a_{n+1}^{y^{p}}\right\}}{\text {COUNT }\left\{i \in\{1, ..., n\}: y_{i}=y^{p}\right\}}
\label{eq:cpc}
\end{equation}
This counts the proportion of calibration samples with the predicted label $y^p$ whose nonconformity scores are $\geq$ than the test sample's score. A low p-value (near $1/n$) suggests high nonconformity, meaning the test sample is significantly different from the training samples. A high p-value (close to 1) indicates strong conformity, showing the test sample closely matches the calibration samples for label $y^p$.

\cparagraph{P-value for regression\label{sec:p_reg}}
We extend classification p-values to regression tasks by generating labels in the calibration dataset using K-means clustering~\cite{arthur2007k}. Specifically, we partition the calibration set into $K$ clusters, $y_1, y_2, \dots, y_K$, based on the feature vectors of each sample. The optimal number of clusters ($K$) is determined using the Gap statistic method~\cite{tibshirani2001estimating}, which compares the within-cluster sum of squares from K-means to that of random clustering over $K$ values from 2 to 20. The Euclidean distance between feature vectors is used as the clustering metric. A larger gap indicates better clustering quality, and \SystemName selects the $K$ with the highest gap.
During deployment, test sample labels are assigned based on the nearest neighbour in the feature space. \SystemName then computes the \emph{p-value}, as in classification, using Equation~\ref{eq:cpc} during both design and deployment.


\subsection{Initialization Assessment\label{sec:dta}}
\SystemName provides a Python function to evaluate whether the framework is properly initialized at design time after obtaining the calibration dataset (Sec.~\ref{sec:mdp}) and the trained underlying model. This is achieved by computing the coverage rate by performing cross-validation on the holdout calibration dataset. Specifically, \SystemName automatically splits the calibration dataset $R$ times ($R=3$ by default) into two \emph{internal} datasets for calibration (80\%) and validation (20\%). It then applies the trained model to the internal validation set and calculates the coverage as:
\begin{equation}
\footnotesize
\frac{1}{R} \sum_{j=1}^{R} \frac{1}{n_{\text{val}}} \sum_{i=1}^{n_{\text{val}}} \mathbb{1} \left\{ y_{i}^{(\text{val})} \in C \left( x_{i}^{(\text{val})} \right) \right\} \approx 1 - \varepsilon
\label{eq:cal}
\end{equation}
where $n_{\text{val}}$ is the size of the validation set, $y_{i}^{(\text{val})}$ is the ground truth of the $i$th validation
example, and $C(x_{i}^{(\text{val})})$ is the prediction region of the $i$th validation example computed by \SystemName using the calibration data. The coverage ratio should be approximately the pre-defined significant level, $1 - \varepsilon$, with minor fluctuations in deviation~\cite{angelopoulos2021gentle}.
A large deviation indicates an ineffective initialization, which usually stems from a poorly trained or designed underlying model. In this case, \SystemName will alert the users when the deviation is more than 0.1, enabling them to enhance the underlying model or adjust the significance level during the design time.

A parameter selection function with a grid search algorithm is provided to help users set the optimal parameters automatically, such as the significant level and cluster size (Sec.~\ref{sec:p_reg}). After evaluating the candidate parameters on the validation dataset, \SystemName will save the selected parameters and use them to predict the confidence in the underlying model at deployment time.

\subsection{Credibilty and Confidence Evaluation\label{sec:cceval}}
\cparagraph{Credibility score}
For each nonconformity function, we use the p-value (Sec.~\ref{sec:nm}) computed for the predicted class as the credibility score. The higher the p-value is, the more likely the test sample is similar to the training-time samples, hence a higher credibility score.

\cparagraph{Confidence score} 
\SystemName estimates the confidence score by evaluating the statistical significance of the prediction using a Gaussian function, $f(x) = e^{-\frac{(x-1)^2}{2 \times c^2}}$, where $c$ (default 3) is a constant, and $x$ is the \emph{prediction set size} for the test sample. The prediction set includes labels likely associated with the test sample, where the nonconformity score exceeds the significance level, $1-\varepsilon$. An empty set suggests the test sample is not linked to any known class, while multiple labels indicate uncertainty, resulting in a low confidence score. As with the credibility score, the prediction set is built from the p-value (Sec.~\ref{sec:nm}). Regression tasks apply the same approach, using the labels introduced by clustering (Sec.~\ref{sec:cpv}). According to our \emph{prediction with rejections} strategy, a sample is flagged as drifting if both scores fall below the significance level.


\subsection{Improve Deployment Time Performance\label{section:IL}}
\SystemName can enhance the performance of deployed ML systems through incremental learning~\cite{gepperth2016incremental, ade2013methods}. For example, suppose a predicted compiler option is likely to be sub-optimal. In that case, the compiler system can use auto-tuning to sample a larger set of configurations to find the optimal one. The idea is to apply other (potentially more expensive) measures to drifting samples. The ground truths can then be added back to the training dataset of the underlying model in a feedback loop for \emph{offline} retraining. Since model retraining occurs only during instances of data drift, it reduces the overhead associated with the collection of training data.

As we will show later, updating a trained model with up to 5\% of identified drifting samples significantly enhances robustness post-deployment. The goal is not to reduce training time but to provide a framework for assessing robustness. Without such a system, frequent retraining or risking performance degradation is required. In code optimization tasks, the main expense is labeling data, not training, and by focusing only on mispredicted samples (e.g. for relabeling), our approach reduces labeling overhead and shortens retraining time.
By filtering out mispredictions, \SystemName detects ageing models and supports implementing corrective methods. This, in turn, will improve user experience and trust in ML systems.


\section{EXPERIMENTAL SETUP\label{sec:setup}}
\begin{table}[!t]
  \scriptsize
  \caption{Case studies and their setups}
  \label{tab:data_summary}
  \begin{tabular}{p{2.6cm}p{2.5cm}p{1.15cm}p{0.5cm}}

        \toprule
        \textbf{\#Use cases} & \textbf{\#Test methods} & \textbf{Models} & \textbf{Tasks}\\
        \midrule
        \rowcolor{Gray} & Magni \etal \cite{magni2014automatic} & MLP  &\\
        \rowcolor{Gray} & \DeepTune~\cite{cummins2017end} & LSTM  &\\
        \rowcolor{Gray} \multirow{-3}{*}{C1: Thread Coarsening} & \IR~\cite{ir2vec} & GBC~\cite{gbc}   &\multirow{-3}{*}{Class.}\\
        & \KStock~\cite{KStock} &SVM &\\
        & \DeepTune~\cite{cummins2017end} &  LSTM &\\
        \multirow{-3}{*}{C2: Loop Vectorization} & Magni \etal & MLP &\multirow{-3}{*}{Class.}\\
        \rowcolor{Gray}& \DeepTune~\cite{cummins2017end}& LSTM& \\
        \rowcolor{Gray}& \Programl~\cite{cummins2021programl}& GNN &\\
        \rowcolor{Gray}\multirow{-3}{*}{C3: Heterogeneous Mapping}& \IR~\cite{ir2vec} & GBC &\multirow{-3}{*}{Class.} \\
         & \Vuldeepecker~\cite{livuldeepecker} & Bi-LSTM  &\\
         & \CodeXGLUE~\cite{lu2021codexglue} &  &\\
         \multirow{-3}{*}{C4: Vulnerability Detection} &  \LineVul~\cite{fu2022linevul} & \multirow{-2}{*}{Transformer}  &\multirow{-3}{*}{Class.}\\
        \rowcolor{Gray} C5: DNN Code Generation & \tlp~\cite{zhai2023tlp} & BERT~\cite{devlin2018bert}   &Reg.\\
        \bottomrule
    \end{tabular}

\end{table}

\cparagraph{Evaluation methodology} As shown in Table \ref{tab:data_summary}, we apply \SystemName to detect drifting samples across 5 case studies, covering 13 representative ML models for classification and regression. We faithfully reproduced all methods following the methodologies in their source publications and used available open-source code. We adhered to the original training methods to ensure \emph{comparable design-time results}.

\cparagraph{Introduce data drift} We introduce changes by separating the training and testing data. We try to mimic practical scenarios by testing the trained model on a benchmark suite not used in the training data or code samples newer than the model training data and the \SystemName calibration dataset. Note that our primary goal is to detect whether \SystemName can successfully detect drifting samples, not to improve the design of the underlying model.

\cparagraph{Prior practices} Prior work often assumes an ideal scenario by splitting training and test samples at the benchmark or method level, where both sets may share similar characteristics~\cite{arp2022and}. In contrast, our evaluation introduces data drift to reflect real-world scenarios where workload characteristics change during deployment. As a result, baseline ML models perform worse on test samples than reported in their original publications~\cite{DBLP:conf/pldi/GoensBECLC19, brauckmann2020compiler}.

\subsection{Case Study 1: Thread Coarsening\label{sec:setupc2}}
This problem develops a model to determine the optimal OpenCL GPU thread coarsening factor for performance optimization. Following ~\cite{magni2014automatic}, an ML model predicts a coarsening factor (ranging from 1 to 32) for a test OpenCL kernel, where 1 indicates no coarsening.

\cparagraph{Underlying models} We consider three ML models designed for this problem: a Multilayer Perceptron (MLP) used in \cite{magni2014automatic}, a long-short-term memory (LSTM) used in \DeepTune~\cite{cummins2017end}, and a Gradient boosting classifier (GBC) used in \IR~\cite{ir2vec}. Like these works, we train and test the models using the labeled dataset from~\cite{magni2014automatic}, comprising 17 OpenCL kernels from three benchmark suites on four GPU platforms.

\cparagraph{Methodology}
As in~\cite{cummins2017end, POEM}, we train the baseline model using leave-one-out cross-validation, which involves training the baseline model on 16 OpenCL kernels and testing on another one. We then repeat this process until all benchmark suites have been tested once.
To introduce data drift, we train the ML models on OpenCL benchmarks from two suites and then test the trained model on another benchmark suite.

\subsection{Case Study 2: Loop Vectorization}
This task constructs a predictive model to determine the optimal Vectorization Factor (VF) and Interleaving Factor (IF) for individual vectorizable loops in C programs~\cite{nuzman2006auto, haj2020neurovectorizer}. Following \cite{haj2020neurovectorizer}, we explore 35 combinations of VF (1, 2, 4, 8, 16, 32, 64) and IF (1, 2, 4, 8, 16). We use LLVM version 17.0 as our compiler, configuring VF and IF individually for each loop using Clang vectorization directives.

\cparagraph{Underlying models}
 We replicate three ML approaches: \KStock~\cite{KStock} (using SVM), \DeepTune~\cite{cummins2017end}, and Magni \etal~\cite{magni2014automatic}, which use neural networks. We use the 6,000 synthetic loops from~\cite{haj2020neurovectorizer}, created by changing the names of the parameters from 18 original benchmarks in the LLVM vectorization test suite. We used the labeled data from~\cite{POEM}, collected on a multi-core system with a 3.6 GHz AMD Ryzen9 5900X CPU and 64GB of RAM.

\cparagraph{Methodology}
Following~\cite{POEM}, we initially allocate 80\% (4,800) of loop programs to train the model, reserving the remaining 20\% (1,200) for testing its performance.
To introduce data drift, we use programs generated from 14 benchmarks for training and evaluate the model on the programs from the remaining 4 benchmarks. 

\subsection{Case Study 3: Heterogeneous Mapping\label{sec:dev_map}}
This task develops a binary classifier to determine if the CPU or the GPU gives faster performance for an OpenCL kernel.

\cparagraph{Underlying models}
We replicated three deep neural networks (DNNs) proposed for this task: \DeepTune~\cite{cummins2017end},\\ \Programl~\cite{cummins2021programl}, and \IR~\cite{ir2vec}. We use the \DeepTune dataset, comprising 680 labeled instances collected by profiling 256 OpenCL kernels from 7 benchmark suites.

\cparagraph{Methodology}
Following~\cite {cummins2017end}, we train and evaluate the baseline model using 10-fold cross-validation. This involves training a model on programs from all but one of the subsets and then testing it on the programs from the remaining subset.
To introduce data drift, we train the models using 6 benchmark suites and then test the trained models on the remaining suites. We repeat this process until all benchmark suites have been tested at least once.

\subsection{Case Study 4: Vulnerability Detection}\label{sec:vul_det}
This task develops an ML classifier to predict if a given C function contains a potential code vulnerability. Following~\cite{Wang2020FUNDED}, we consider the top-8 types of bugs from the 2023 CWE~\cite{cwe25}.

\cparagraph{Underlying models} We replicated four representative ML models designed for bug detection: \CodeXGLUE~\cite{lu2021codexglue}, \LineVul~\cite{fu2022linevul}, both based on Transformer networks, and \Vuldeepecker\\~\cite{livuldeepecker} which based on a Bi-LSTM network. We evaluate this task with a dataset comprising 4,000 vulnerable C program samples labeled with one of the eight vulnerability types, each with around 500 samples. The vulnerable code samples cover 2013 and 2023 and are collected from the National Vulnerability Database (NVD), CVE, and open datasets from GitHub.

\cparagraph{Methodology}
As with prior approaches, we initially train the model on 80\% of the randomly selected samples and evaluate its performance on the remaining 20\% samples.
Then, we introduce data drift by training the model on data collected between 2013 and 2020 and testing the trained model on samples collected between 2021 and 2023.

\subsection{Case Study 5: DNN Code Generation\label{sec:tvm}}
This task builds a \emph{regression-based} cost model to drive the schedule search process in TVM~\cite{chen2018tvm} for DNN code generation on multi-core CPUs. The cost model estimates the potential gain of a schedule (e.g., instruction orders and data placement)  to guide the search.

\cparagraph{Underlying model}
We apply \SystemName to \tlp~\cite{zhai2023tlp}, a cost model-based tensor program tuning method integrated into the TVM compiler v0.8~\cite{chen2018tvm}. We use $2,308 \times 4$ samples collected from 4 Transformer-based BERT models of different sizes in the TenSet dataset~\cite{zheng2021tenset}.

\cparagraph{Methodology}
For the baseline, we trained and tested the cost model on the BERT-base dataset, where the model is trained on 80\% (400K) randomly selected samples and then tested on the remaining 20\% (100K)samples. To introduce data drift, we tested the trained model on the other three variants of the BERT model. We ran the TVM search engine for around 8 hours (4,000 iterations) for each DNN on a 12-core 2.7GHz AMD EPYC 9B14 CPU server.

\subsection{Performance Metrics\label{sec:pr}}

\cparagraph{Performance to the oracle} 
For code optimization tasks (case studies 1 to 3), we compute the ratio of the predicted performance to the best performance obtained by exhaustively trying all options. A ratio of 1.0 means the prediction leads to the best performance given by an ``oracle" method.

\cparagraph{Misprediction threholds} 
For code optimization, we consider a prediction to be a misprediction if runtime performance is 20\% or more below the Oracle performance (case studies 1–3) or if predicted performance deviates by 20\% or more from profiling results (case study 5). For bug detection (case study 4), a misprediction happens when the model misclassifies a test input.

\bparagraph{Coverage deviation}
This ``\emph{smaller-is-better}'' metric measures the difference between the confidence level and \SystemName's true coverage rate on the model. A zero deviation means the coverage rate matches the predefined significance level.

\cparagraph{Metrics for data drift detection} We consider the following ``\emph{higher-is-better}'' metrics for detecting drifting samples:

\bparagraph{Accuracy} The ratio of the number of correctly predicted samples to the total number of testing samples.

\bparagraph{Precision} The ratio of mispredicted samples that were correctly rejected to the total number of mispredicted samples. This metric answers questions like ``\emph{Of all the rejected predictions, how many are actually mispredictions?}''. High precision indicates a low \emph{false-positive} rate, meaning \SystemName rarely incorrectly rejects predictions.

\bparagraph{Recall (or Sensitivity)} The ratio of mispredicted samples that were correctly rejected to the total number of mispredicted samples. This metric answers questions like ``\emph{Of all the mispredicted test samples, how many are rejected by \SystemName?}''. High recall suggests a low \emph{false-negative} rate, indicating \SystemName can identify most mispredictions.

\bparagraph{F1 score} The harmonic mean of Precision and Recall, calculated as $2\times\frac{Recall \times Precision} {Recall +
Precision}$. It is useful when the test data has an uneven distribution of drifting and normal samples. The highest possible F1 score is 1.0, indicating perfect precision and recall.

\begin{table}[t]
    \centering
    \caption{Summary of our main evaluation results}
    \label{tab:result_summary}
    \scriptsize
    \begin{tabularx}{0.45\textwidth}{p{0.9cm}p{0.9cm}p{1.8cm}p{0.5cm}p{0.5cm}p{0.5cm}p{0.5cm}}
        \toprule
        \multicolumn{3}{c}{\textbf{Perf. to the Oracle}} & \multicolumn{4}{c}{\textbf{\SystemName performence}}\\
        \rowcolor{Gray}\textbf{Training} & \textbf{Deploy.} & \textbf{\SystemName on deploy.} & \textbf{Acc.} & \textbf{Pre.} & \textbf{Recall} & \textbf{F1} \\
        \midrule
          0.836 & 0.544 & 0.807 & 86.8\% & 86.0\% & 96.2\% & 90.8\% \\
        \bottomrule
    \end{tabularx}
\end{table}

\begin{figure*}[t]
           \subfigure[C1: thread coarsening]{
    \begin{minipage}[t]{0.23\textwidth}
    \centering
    \includegraphics[width=\columnwidth]{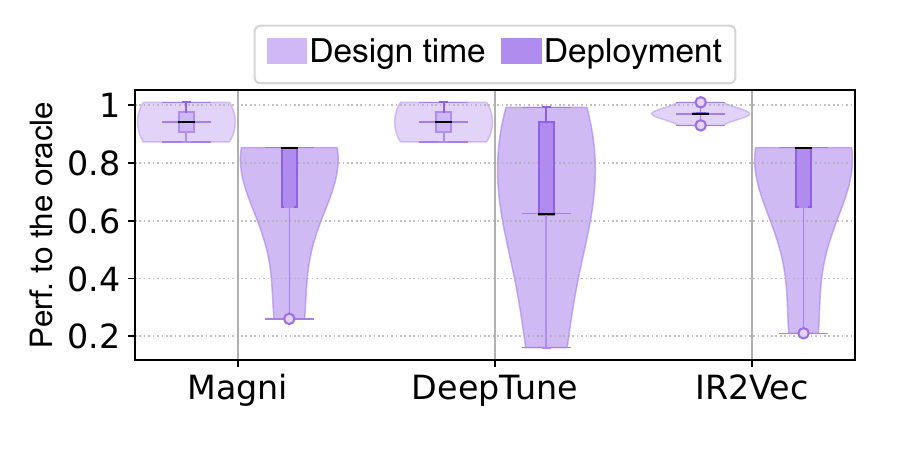}
    \end{minipage}\label{fig:deploy_case1}
    }
    \subfigure[C2: loop vectorization]{
    \begin{minipage}[t]{0.23\textwidth}
    \centering
    \includegraphics[width=\columnwidth]{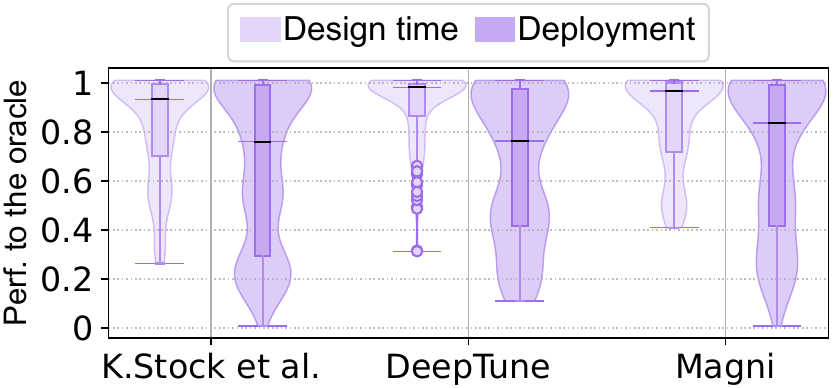}
    \end{minipage}\label{fig:deploy_case2}
    }
    \subfigure[C3: heterogeneous device mapping]{
    \begin{minipage}[t]{0.23\textwidth}
    \centering
    \includegraphics[width=\columnwidth]{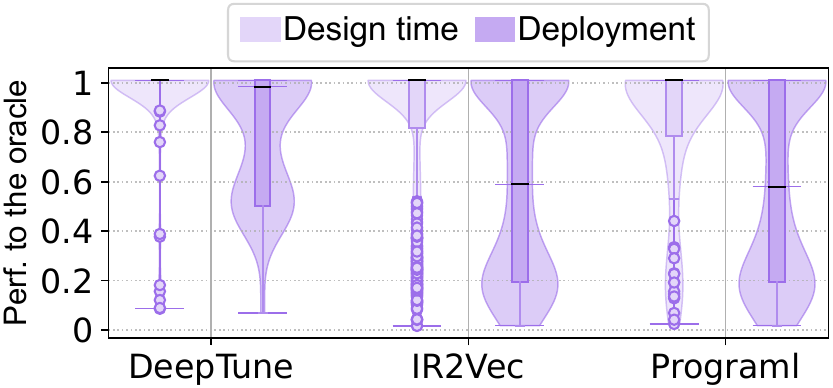}
    \end{minipage}\label{fig:deploy_case3}
    }
    \subfigure[C4: vulnerability detection]{
    \begin{minipage}[t]{0.23\textwidth}
    \centering
    \includegraphics[width=\columnwidth]{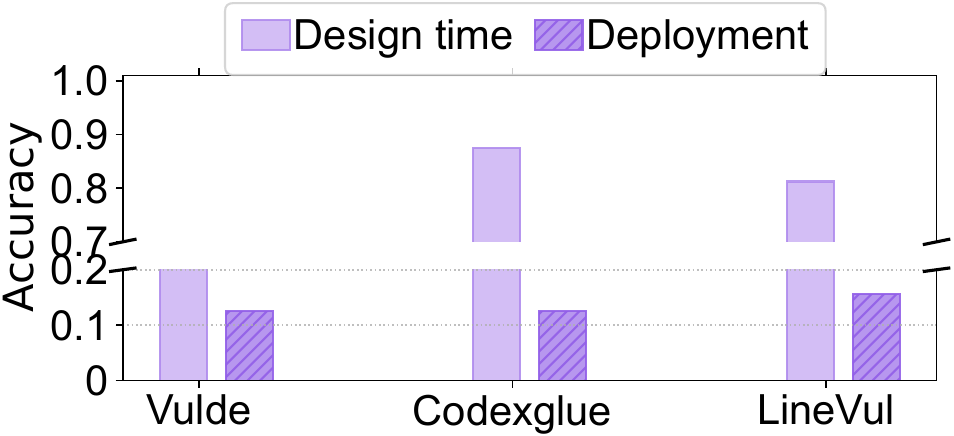}
    \end{minipage}\label{fig:deploy_case4}
    }
    \centering
     \caption{The resulting performance when using an ML model for decision making.
     The performance of all learning-based models can suffer during the deployment phase when the test samples significantly differ from the training data.
     }\label{fig:drifting}

\end{figure*}

\begin{figure}[t]
           \subfigure[C1: thread coarsening]{
    \begin{minipage}[t]{0.22\textwidth}
    \centering
    \includegraphics[width=\columnwidth]{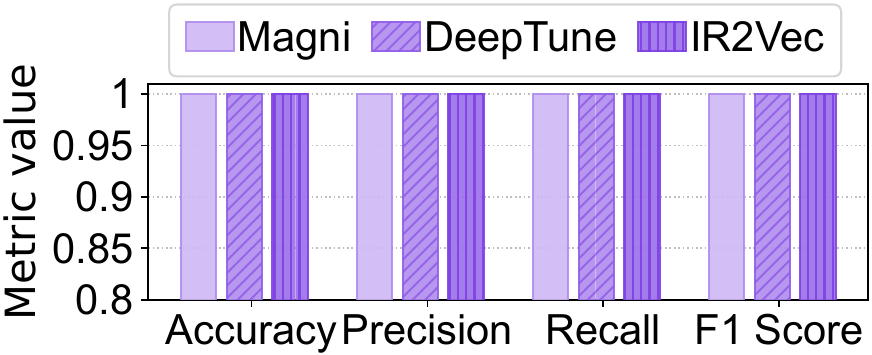}

    \end{minipage}\label{fig:drift_case1}
    }
    \subfigure[C2: loop vectorization]{
    \begin{minipage}[t]{0.22\textwidth}
    \centering
    \includegraphics[width=\columnwidth]{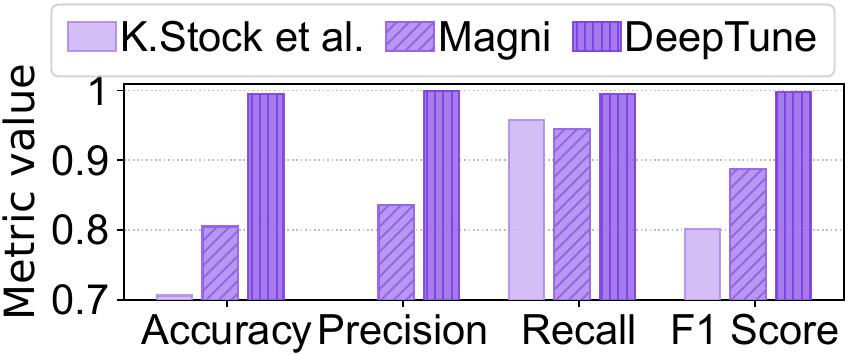}
    \end{minipage}\label{fig:drift_case2}
    }

        \subfigure[C3: heterogeneous mapping]{
    \begin{minipage}[t]{0.22\textwidth}
    \centering
    \includegraphics[width=\columnwidth]{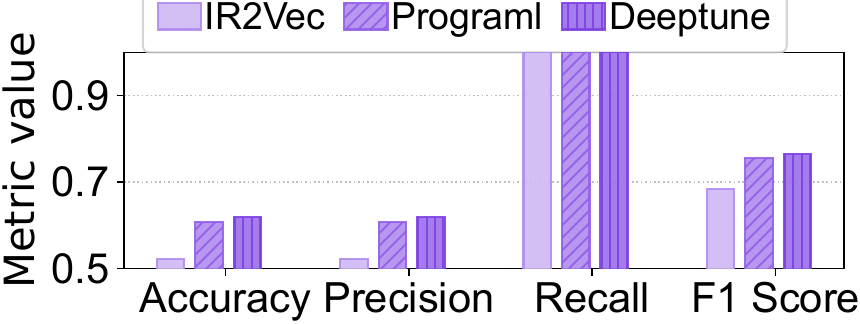}
    \end{minipage}\label{fig:drift_case3}
    }
    \subfigure[C4: vulnerability detection]{
    \begin{minipage}[t]{0.22\textwidth}
    \centering
    \includegraphics[width=\columnwidth]{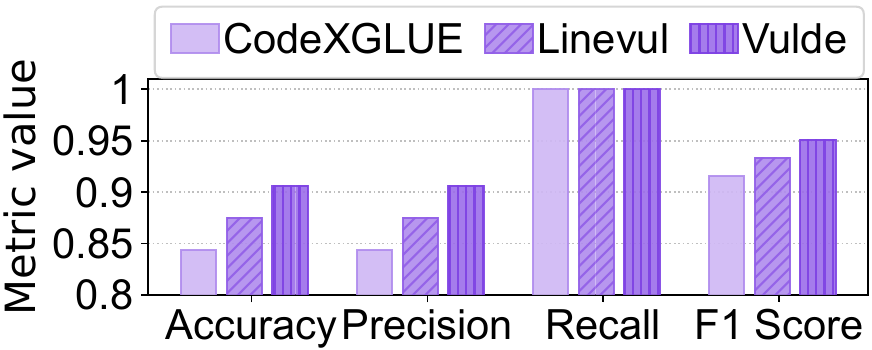}
    \end{minipage}\label{fig:drift_case4}
    }
    \subfigure[C5: DNN code generation]{
    \begin{minipage}[t]{0.25\textwidth}
    \includegraphics[width=\columnwidth]{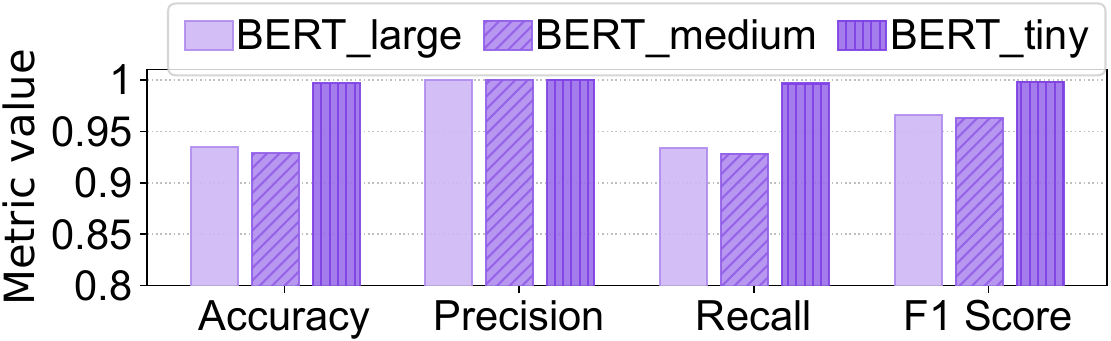}
    \end{minipage}\label{fig:drift_case5}
    }
    \centering
    \caption{\SystemName's performance for detecting drifting samples across case studies and underlying models (higher is better).}
\label{fig:detect_drift}
\end{figure}

\begin{figure*}[t]
           \subfigure[C1: thread coarsening]{
    \begin{minipage}[t]{0.23\textwidth}
    \centering
    \includegraphics[width=\columnwidth]{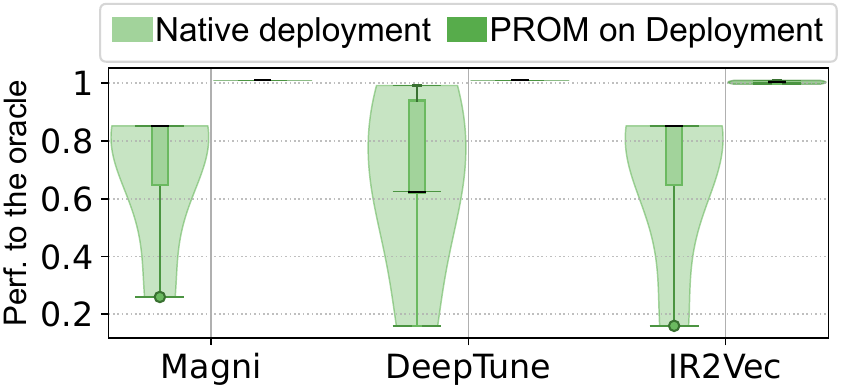}

    \end{minipage}\label{fig:IL_case1}
    }
    \subfigure[C2: loop vectorization]{
    \begin{minipage}[t]{0.23\textwidth}
    \centering
    \includegraphics[width=\columnwidth]{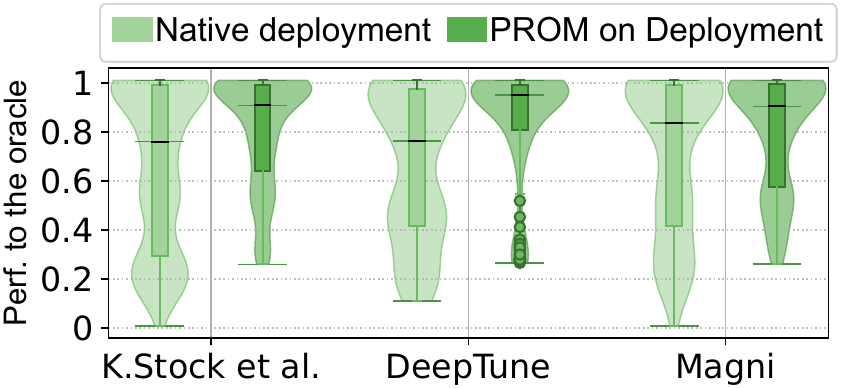}
    \end{minipage}\label{fig:IL_case2}
    }
    \subfigure[C3: heterog. mapping]{
    \begin{minipage}[t]{0.23\textwidth}
    \centering
    \includegraphics[width=\columnwidth]{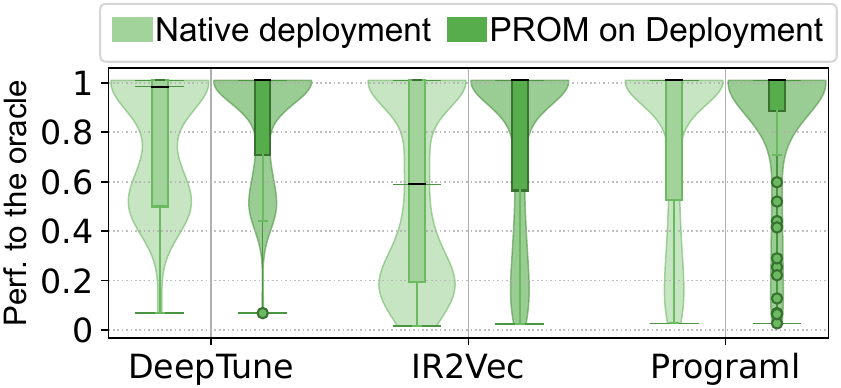}
    \end{minipage}\label{fig:IL_case3}
    }
    \subfigure[C4: vuln. detection]{
    \begin{minipage}[t]{0.23\textwidth}
    \centering
    \includegraphics[width=\columnwidth]{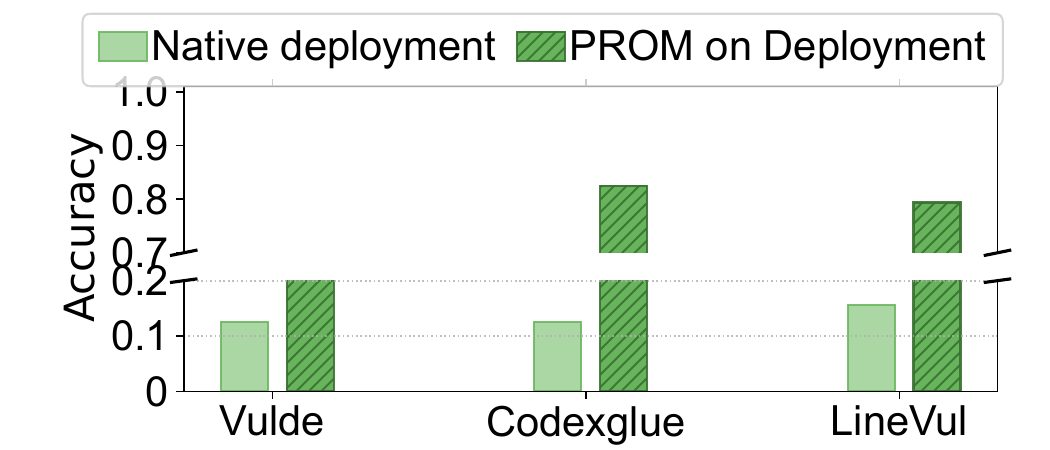}
    \end{minipage}\label{fig:IL_case4}
    }
    \centering
    \caption{\SystemName enhances performance through incremental learning in different underlying models.}\label{fig:IL_drift}
\end{figure*}

\begin{figure}[!t]
    \centering
    \includegraphics[width=0.35\textwidth]{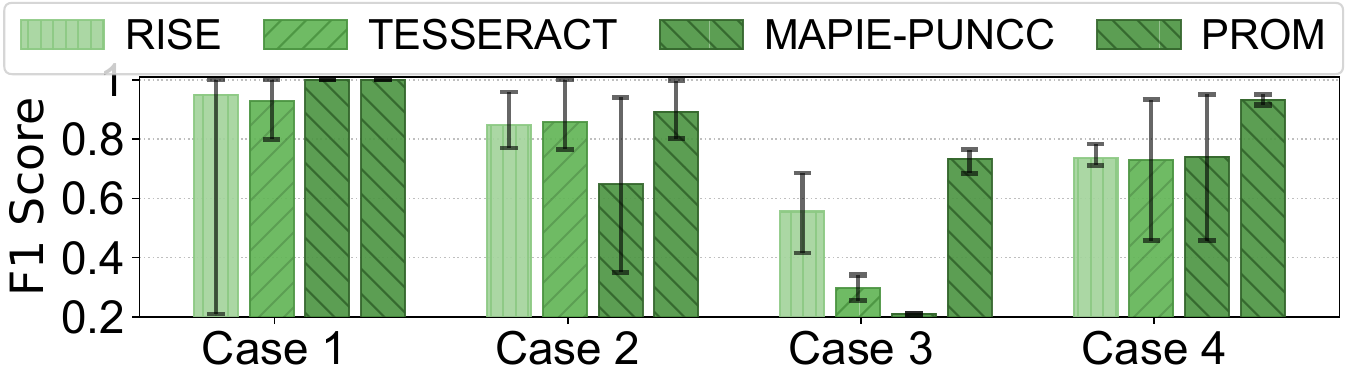}\\
    \caption{Geometric mean and variances across models of the F1 score in classification tasks.}
    \label{fig:significance}
\end{figure}

\section{EXPERIMENTAL RESULTS}\label{section:results}
\cparagraph{Highlights} Table~\ref{tab:result_summary} summarizes the main results of our evaluation. All the tested models were impacted by changes in the application workloads (Sec.~\ref{sec:overall_results}), where the performance relative to the Oracle predictor drops significantly from training time to deployment time. \SystemName can detect 96.2\% of the drifting samples on average (Sec.~\ref{sec:detect}). When combined with incremental learning, \SystemName enhances the performance of deployed models, improving prediction performance by up to 6.6x (Sec.~\ref{sec:improve}). \SystemName also outperforms existing CP-based methods and related work (Sec.~\ref{sec:compare}). 


\subsection{Impact of Drifting Data}\label{sec:overall_results}

This experiment assesses the impact of data drift by applying a trained model to programs from an unseen benchmark suite or CVE dataset (Sec.~\ref{sec:setup}). For code optimization tasks (case studies 1-3 and 5), we report the performance ratio relative to the oracle (Sec.~\ref{sec:pr}). In case study 4, we report the accuracy of bug prediction.

Figure~\ref{fig:drifting} shows the classifiers' performance (case studies 1-4) during design and deployment, while Table~\ref{tab:reg} presents the regression model's performance (case study 5). The violin diagrams in Figure~\ref{fig:drifting} show the distribution of test sample performance, with the violin's width representing the number of samples in each range. The line inside shows the median, and the top and bottom lines indicate the extremes. Outliers are marked as circles. Ideally, a model's violin would be wide at the top, reflecting good performance for most samples.

\cparagraph{Design time performance}  
For case studies 1-4, we assess design-time performance by holding out 10\% of the training samples as a validation set and applying the trained model to it. In case study 5, the training data covers all DNN models, but the model is tested on samples from unseen schedules. This process is repeated 10 times, and the average performance is used as the design-time result. This \emph{ideal setting} assumes that validation and training samples come from the same project or benchmark, with similar workload patterns, yielding comparable results to those reported in the original model publications.


\cparagraph{Deployment time performance} An ML model's robustness can suffer during deployment. From  Figure~\ref{fig:drifting}, this can be observed from the bimodal distribution of the violin shape or a lower prediction accuracy at the deployment stage. From the violin diagrams, we observe a wider violin shape towards the bottom of the y-axis and a lower median value compared to the design-time result. From Table~\ref{tab:reg}, this also can be seen from a lower deployment-time prediction accuracy than the design-time performance. The impact of drifting samples is clearly shown in vulnerability detection of Figure~\ref{fig:deploy_case4}, where the prediction accuracy drops by an average of 62.5\%. For DNN code generation (Table~\ref{tab:reg}), the accuracy of performance estimation can also drop from 84.5\% to as low as 22.4\%. The results highlight the impact of data drift.

\subsection{Detecting Drifting Samples\label{sec:detect}}
Figure~\ref{fig:detect_drift} reports \SystemName's performance in predicting drifting samples across case studies and underlying models. For all tasks, \SystemName achieves an average precision of 0.86 with an average accuracy of 0.87. This means it rarely filters out correct predictions. For the binary classification task of heterogeneous mapping (Figure~\ref{fig:drift_case3}), \SystemName achieves an average F1 score of 0.74. In this case, \SystemName sometimes rejects correct predictions. This is because the probability distribution of binary classification is often less informative for CP than in multiclass cases~\cite{taquet2022mapie}. 
For the regression task of case study 5, \SystemName can detect most of the drifting samples with a recall of 0.95 and an average precision of 1.
Furthermore, the underlying model's quality also limits the performance of \SystemName. When the information given by the underlying model becomes noisy,  \SystemName can be less effective.
Averaged across case studies, in detecting mispredictions, \SystemName achieves a recall of 0.96, a false-positive rate of 0.14 and a false-negative rate of 0.04, suggesting that \SystemName is highly accurate in detecting drifting samples.

\subsection{Incremental Learning}\label{sec:improve}

\begin{table}[!t]
    \centering
    \caption{C5: DNN code generation (performance to the oracle ratio) - trained on BERT-base and tested on BERT variants.}
    \label{tab:reg}
    \scriptsize
    \begin{tabularx}{0.45\textwidth}{p{2.7cm}p{0.9cm}p{0.9cm}p{0.9cm}p{0.9cm}}
        \toprule
        \textbf{Network} &  \textbf{BERT-base} & \textbf{BERT-tiny} & \textbf{BERT-medium} & \textbf{BERT-large}\\
        \midrule
        \rowcolor{Gray}  Native deployment &0.845 & 0.224& 0.668& 0.703 \\
          \SystemName assisted deployment& / & 0.794&  0.810&   0.808     \\
        \bottomrule
    \end{tabularx}
\end{table}
In this experiment, we use \SystemName to identify drifting samples and update the underlying models by retraining with a small set of \SystemName-identified samples. \SystemName preserves the performance of the methods close to their original levels, as shown by improved accuracy (Figure~\ref{fig:IL_case4}), the performance-to-oracle ratio (Table~\ref{tab:reg}), and violin plots (Figures~\ref{fig:IL_case1} to \ref{fig:IL_case3}), where test sample distributions shift towards higher performance with a better median than native deployment without \SystemName.
Overall, \SystemName requires labeling at most 5\% (sometimes just one) of drifting samples to update the model. Without it, one would need to label random test samples, leading to higher maintenance costs and unnecessary user confirmations for samples the model can predict correctly.



\subsection{Individual Case Studies\label{sec:ics}}
We now examine case studies showing how \SystemName improves the underlying model with incremental learning.

\cparagraph{Case study 1: thread coarsening}
In this experiment, we tested the trained model on kernels from OpenCL benchmarks unseen at the model training stage. Figure~\ref{fig:deploy_case1} and~\ref{fig:IL_case1} show that the performance of all ML models drops as the test dataset changes.
By relabeling just one drifting sample using incremental learning, \SystemName improves the performance to the oracle ratio from an average of 77.6\% to 99.0\% (21.4\% improvement) during deployment.

\cparagraph{Case study 2: loop vectorization}
This experiment introduced changes by testing the underlying model on loops extracted from unseen benchmarks. Figure~\ref{fig:deploy_case2} and~\ref{fig:IL_case2} show that drifting data led to a performance reduction for all methods, averaging 9\%.
Retraining the model using only 5\% of the \SystemName-identified drifting samples helps restore the underlying model's initial performance, leaving only a 1.6\% gap to the design-time performance.

\cparagraph{Case study 3: heterogeneous mapping}
This experiment tests the underlying models on OpenCl kernels from \emph{unseen} benchmarks. Figures~\ref{fig:deploy_case3} and \ref{fig:IL_case3} show that all ML models deliver low performance on \emph{unseen} benchmarks. \SystemName also successfully detects 100\% of the drifting samples (recall) on average with an accuracy rate of 58\%. Further, by utilizing incremental learning on 5\% of the drifting samples, \SystemName improved the performance to oracle ratio of all systems from 63.1\% to 78.9\%  (15.8\% improvement) on average.

\cparagraph{Case study 4: vulnerability detection}
In this experiment, we tested a trained model on a vulnerability dataset from a time period not covered by the training data. Figures~\ref{fig:deploy_case4} and~\ref{fig:IL_case4} show that all models initially had low prediction accuracy, ranging from 12.5\% to 15.6\% when facing new code patterns. \SystemName correctly identified all mispredictions with a recall of 1. By relabeling up to 5\% of the \SystemName-identified drifting samples and updating the model, we improved the accuracy from an average of 13.5\% to 72.5\%, achieving 95.3\% of the design-time performance.


\cparagraph{Case study 5: DNN code generation}~\label{sec:regression}
In this experiment, we apply the cost model trained on the Bert-base dataset to three other variants of the BERT network. Once again, from Table~\ref{tab:reg}, we see the performance of the trained model experiences a reduction from 84.5\% to 53.2\%. For BERT-tiny, the performance drops as much as 65.3\%. \SystemName can detect 95.3\% of drifting data and achieve a precision of 1. After profiling just 5\% of the \SystemName-identified drifting data and using them to retrain the cost model online during the TVM code search process, the average performance of the cost model improves to 80.4\%, resulting in a 2.0x enhancement.

\begin{figure*}[t]
           \subfigure[Case study 1: thread coarsening]{
    \begin{minipage}[t]{0.235\textwidth}
    \centering
    \includegraphics[width=\columnwidth]{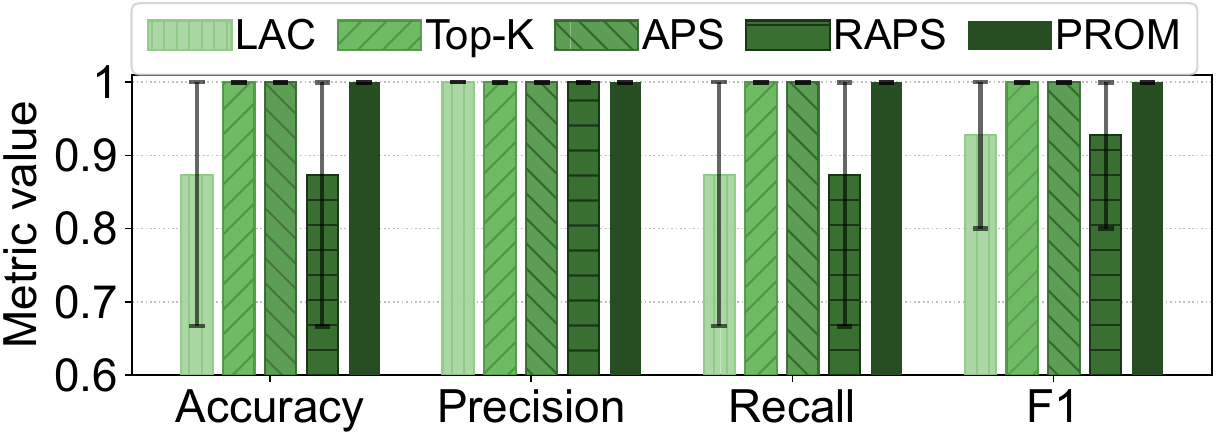}
    \end{minipage}
    }
    \subfigure[Case study 2: loop vectorization]{
    \begin{minipage}[t]{0.235\textwidth}
    \centering
    \includegraphics[width=\columnwidth]{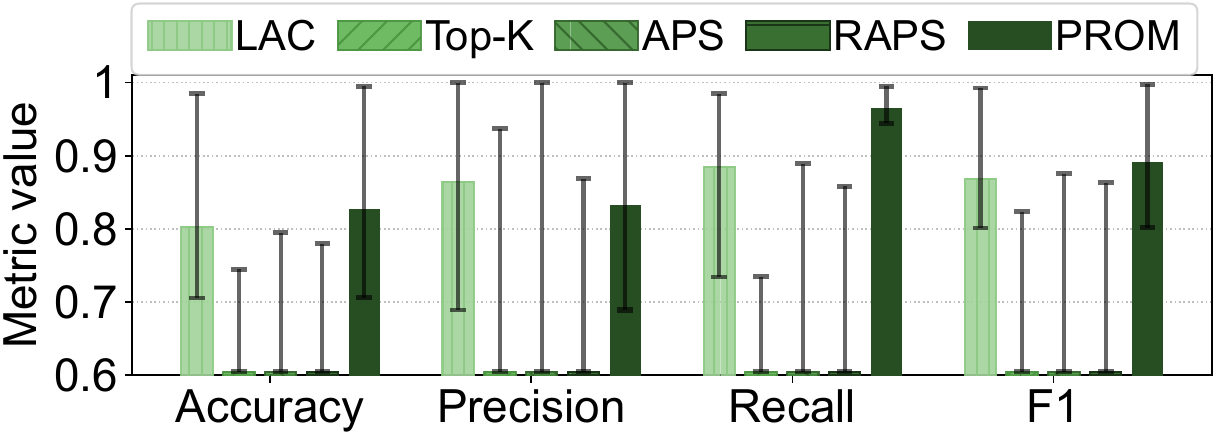}
    \end{minipage}
    }
    \subfigure[Case study 3: heterog. mapping]{
    \begin{minipage}[t]{0.235\textwidth}
    \centering
    \includegraphics[width=\columnwidth]{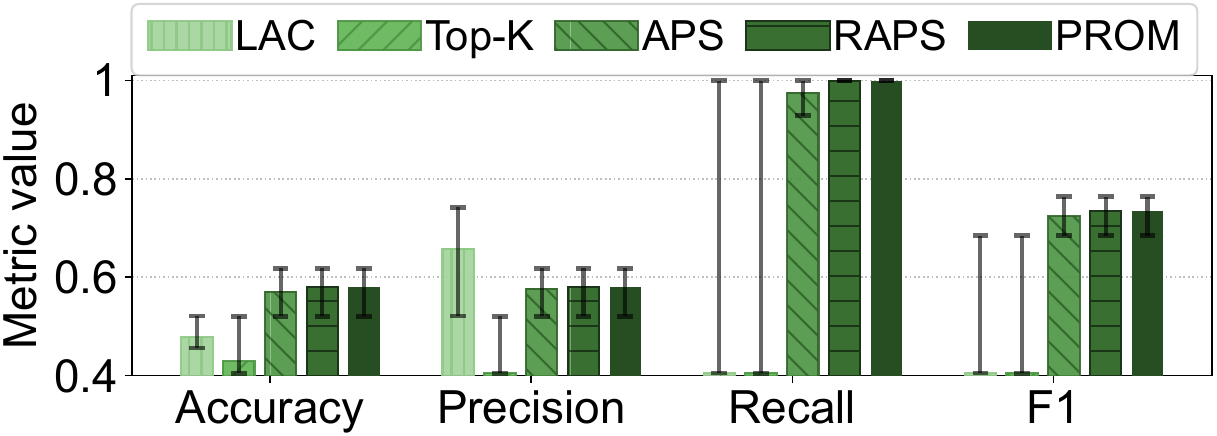}
    \end{minipage}
    }
    \subfigure[Case study 4: vuln. detection]{
    \begin{minipage}[t]{0.235\textwidth}
    \centering
    \includegraphics[width=\columnwidth]{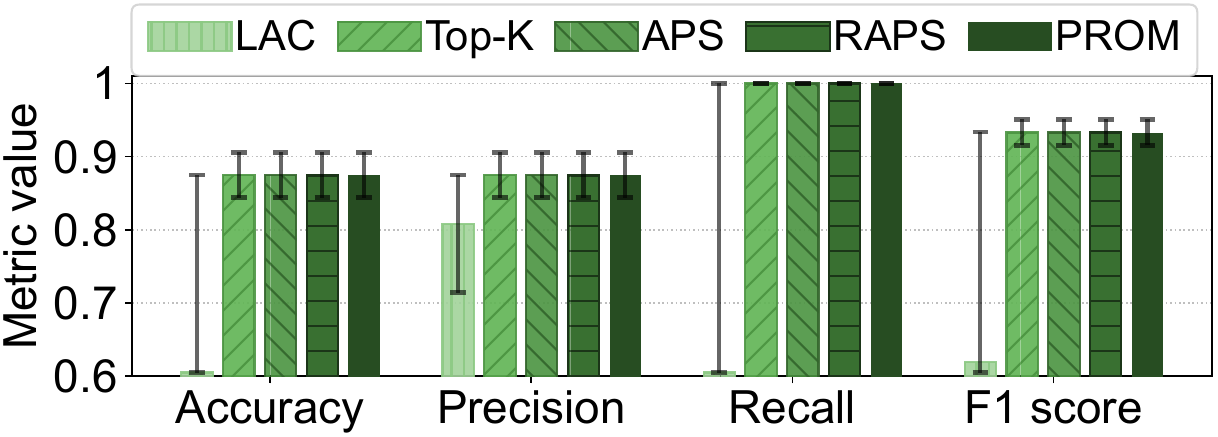}
    \end{minipage}
    }
    \centering
    \caption{Performance of individual nonconformity functions. Min-max bar shows the performance across ML models.}\label{fig:choice}
\end{figure*}


\subsection{Comparison to CP Libraries and Methods~\label{sec:compare}}

We compare \SystemName with \Mapie~\cite{taquet2022mapie} and \Puncc~\cite{mendil2023puncc}, two CP libraries used for outlier detection, as well as \RAISE~\cite{RISE} and \TESSERACT~\cite{sec2019TESSERACT}, which use a single nonconformity function. \RAISE also trains an SVM for misprediction detection but, like \TESSERACT, supports only classical classifiers, so we evaluate them on cases 1 to 4. Figure~\ref{fig:significance} shows average results, with min-max bars indicating variation across models. \SystemName outperforms \TESSERACT by 17.6\%, achieving a higher F1 score due to its improved nonconformity strategy. \RAISE struggles with uneven data or tasks with many labels, while \SystemName's model-free ensemble approach handles these cases better. Naive CP yields the lowest F1 score, showing its limitations in detecting drift in large-scale tasks.


\begin{figure}[!t]
    \centering
    \includegraphics[width=0.35\textwidth]{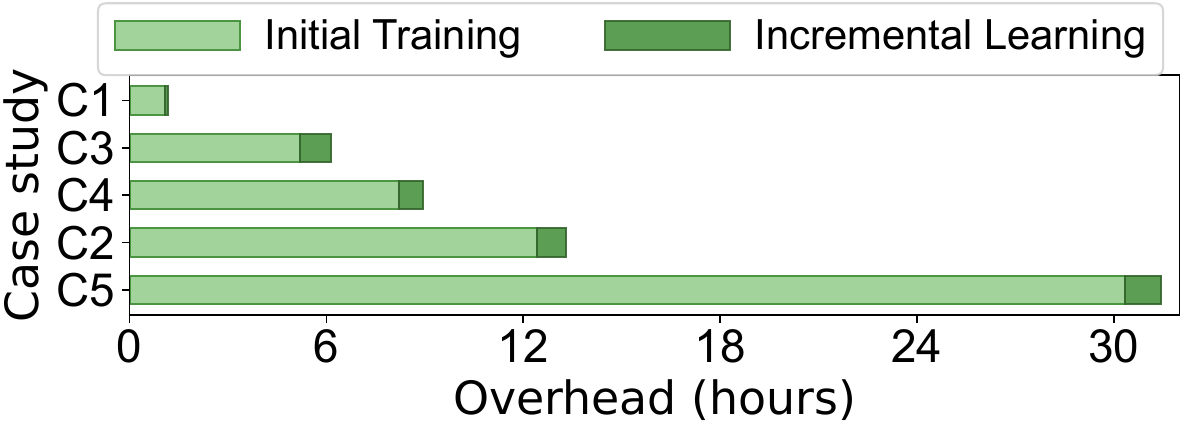}\\
    \caption{The average training and incremental learning overhead of individual case studies.}
    \label{fig:overhead}
\end{figure}

\begin{figure}[t]
               \subfigure[\SystemName performance as the threshold increases in loop vectorization]{
    \begin{minipage}[t]{0.21\textwidth}
    \centering
    \includegraphics[width=\columnwidth]{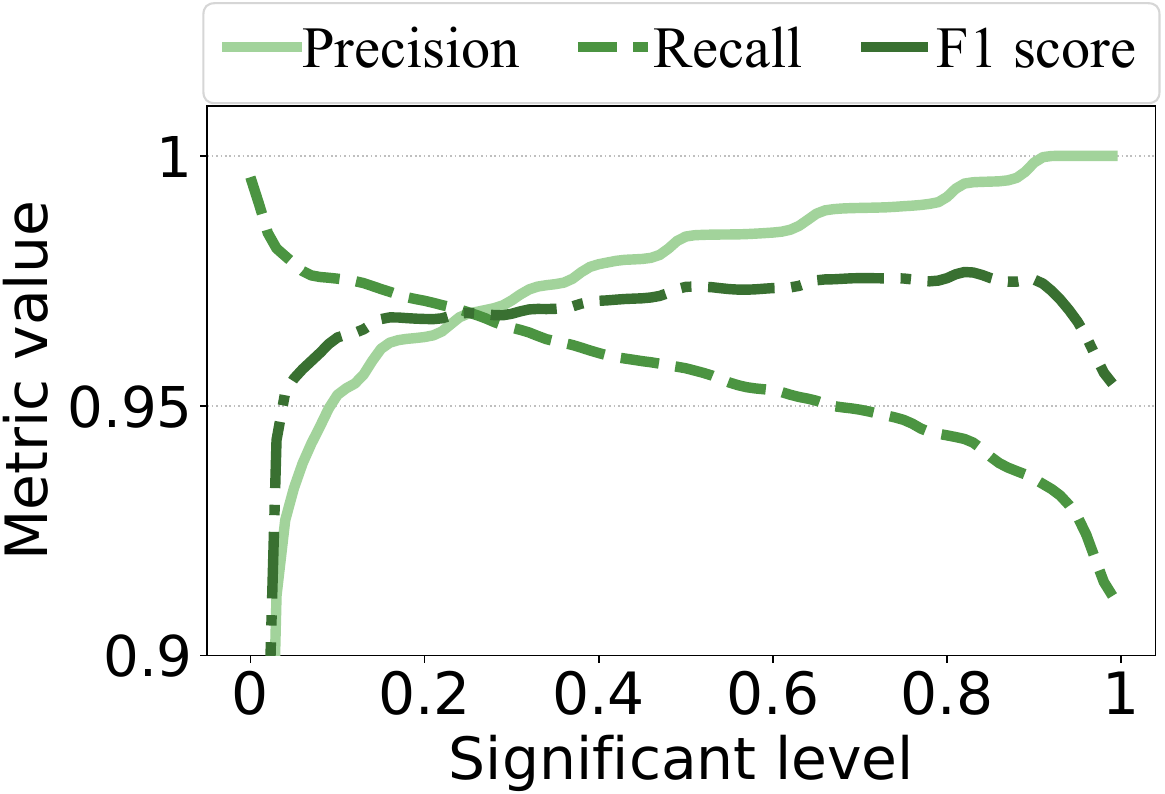}
    \end{minipage}\label{fig:sensi_threshold}
    }
    \vspace{-1mm}
\subfigure[\SystemName performance as cluster size increases in regression task]{
    \begin{minipage}[t]{0.21\textwidth}
    \centering
    \includegraphics[width=\columnwidth]{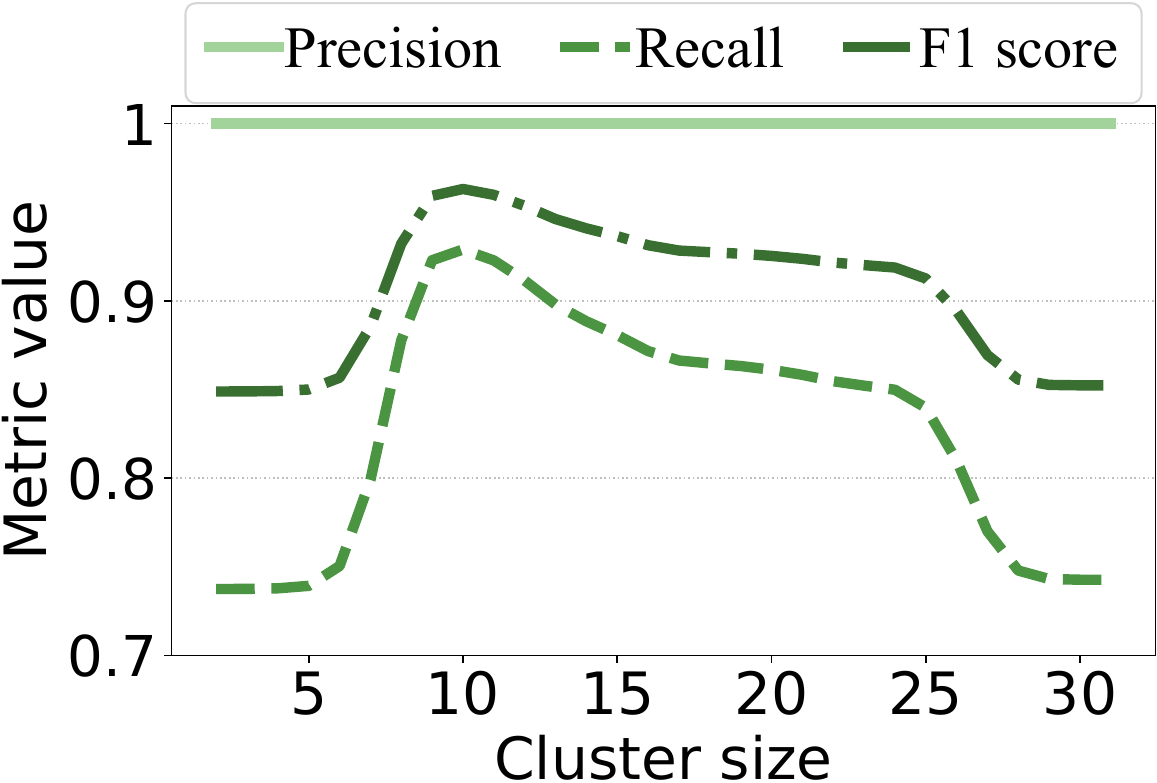}
    \end{minipage}\label{fig:sensi_cluster}
    }

        \subfigure[\SystemName performance as Gaussian scale parameter changes]{
    \begin{minipage}[t]{0.21\textwidth}
    \centering
    \includegraphics[width=\columnwidth]{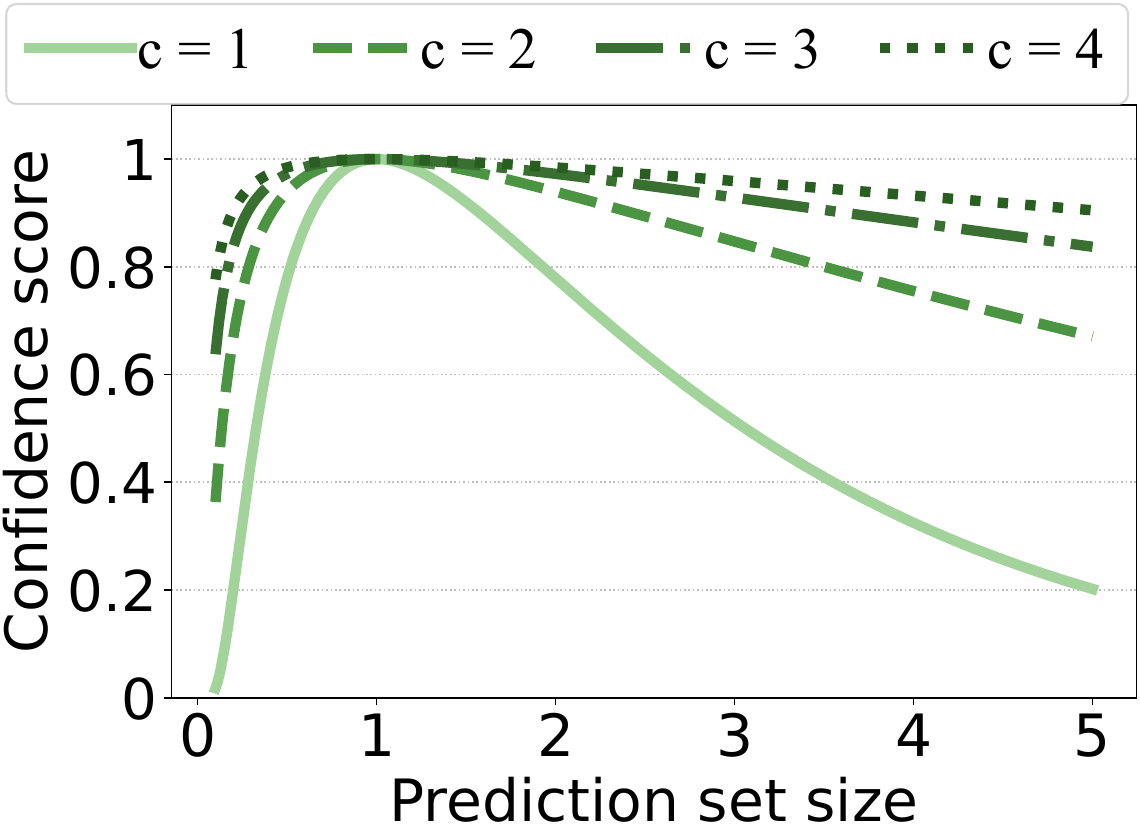}
    \end{minipage}\label{fig:sensi_gaussian}
    }
    \subfigure[Coverage deviations across 5 case studies]{
    \begin{minipage}[t]{0.22\textwidth}
    \centering
    \includegraphics[width=\columnwidth]{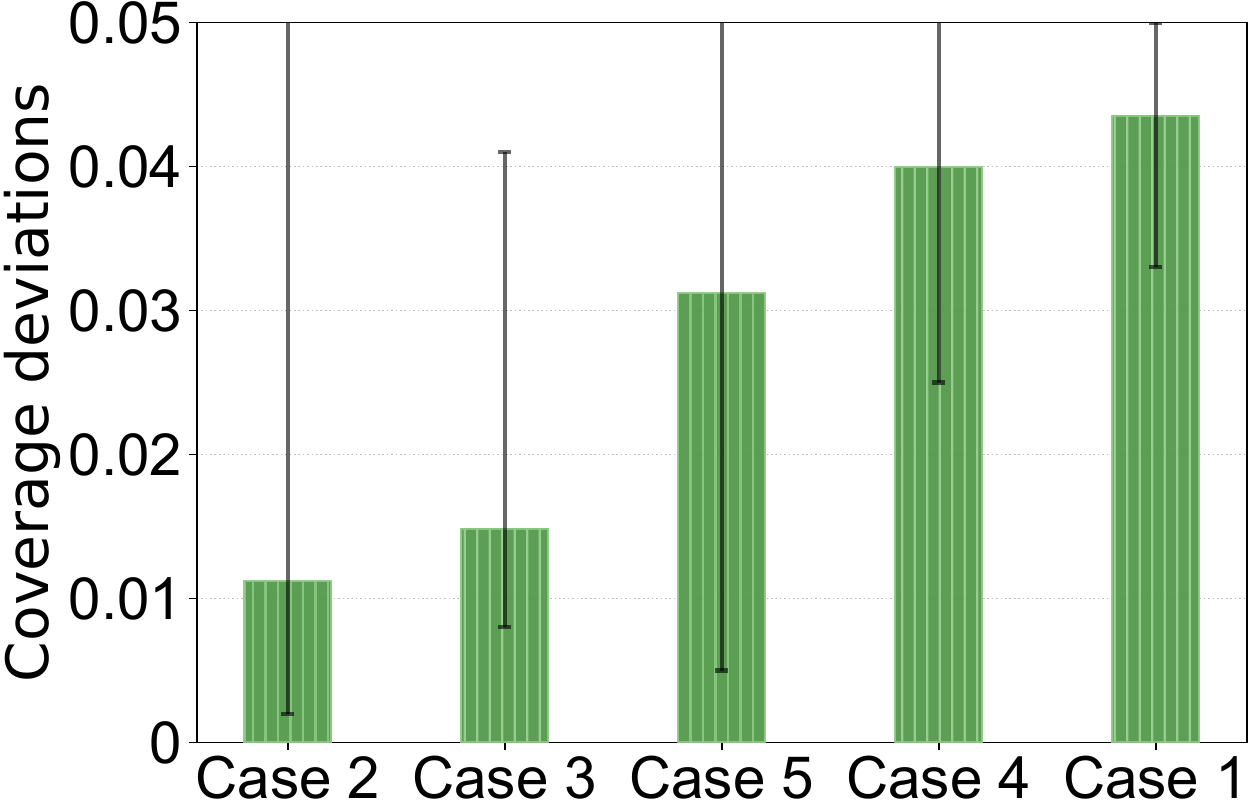}
    \end{minipage}\label{fig:coverage}
    }

    \centering

    \caption{Sensitive analysis of \SystemName hyperparameters.}\label{fig:IL_drift}
\end{figure}

\subsection{Further Analysis\label{sec:fa}}
\cparagraph{Nonconformity functions}~\label{res:choice}
Figure~\ref{fig:choice} shows the average performance of the four default \SystemName\ functions in detecting drifting samples across case studies, with min-max bars indicating variance across models. \SystemName's ensemble strategy outperforms individual functions in all metrics, demonstrating that no single function performs well across all case studies. By combining multiple functions, \SystemName\ enhances the generalization of statistical assessments.


\cparagraph{Sensitive analysis}
Figures~\ref{fig:sensi_threshold} to \ref{fig:sensi_gaussian} show how the significance threshold and other parameters (e.g. cluster size and Gaussian scale) impact \SystemName's performance for data drift detection. A larger significance threshold reduces false positives, improving precision but potentially lowering recall. \SystemName automatically determines the optimal cluster size using the Gap statistic method~\cite{tibshirani2001estimating}. Deviations from the optimal cluster size affect detection performance. For the Gaussian scale parameter, prediction set sizes smaller or larger than 1 lead to reduced confidence by suggesting too few or too many classes for a sample.

\cparagraph{Coverage deviation}
Figure~\ref{fig:coverage} shows the coverage deviation across five cases. The min-max bar represents the variance across the underlying models. The geomean of deviation is 2.5\%, which is a benign fluctuation and indicates a good fit for conformal prediction on the underlying models. The thread coarsening task shows a 4.4\% deviation due to the small calibration dataset, which could be mitigated by adding more calibration samples.

\cparagraph{Training overhead}
Figure~\ref{fig:overhead} reports the overhead of initial model training and incremental learning. Initial training takes several hours to one day, while incremental learning with fine-tuning takes less than one hour on a multi-core CPU or a desktop GPU - a negligible overhead compared to the initial training time. 

\cparagraph{Runtime overhead}
During deployment, the main runtime overhead in \SystemName comes from computing nonconformity scores and detecting drifting samples. This overhead is minimal: on a low-end laptop, computing confidence and credibility scores and performing drift detection take less than 10 and 2 milliseconds, respectively.
\section{DISCUSSIONS\label{section:discussions}}
Naturally, there is room for improvement and further work. We discuss a few points here.

\cparagraph{Overfitting}
To mitigate overfitting, \SystemName retrains the model using the original dataset and a few \SystemName-identified mispredictions. Incorporating additional data from the deployment environment likely enhances the model's generalization. In our experiments, cross-validation shows that this retraining approach effectively improves generalization.

\cparagraph{Robustness}
\SystemName employs a model-free approach to detect drifting samples, avoiding reliance on supervised learning. The calibration dataset can be updated during incremental learning, and our detection framework is updated during ML model retraining. Thus, \SystemName can adapt over time to changes in model training samples.

\cparagraph{Combining with reinforcement learning}
\SystemName is a good fit for RL in code optimization, which usually requires a cost model to estimate the reward to avoid expensive program profiling. \SystemName can assess the RL cost model's robustness, guiding RL to profile only samples likely to be mispredicted by the cost model (similar to case study 5). This profiling information can then be used to update and improve the cost model during the RL search.

\cparagraph{Applicability}
\SystemName targets probability-based classifiers~\cite{dobson2018introduction, dominguez2011logistic, xanthopoulos2013linear, rish2001empirical}, making it broadly applicable across ML algorithms. It can also be used during training to evaluate whether adding new data would improve performance, helping reduce costly data labeling for code optimization~\cite{ogilvie2017minimizing}.


\cparagraph{Rejection costs}
Handling rejected predictions varies by application. Simple cases may involve manual review or other optimization strategies, such as iterative compilation for optimal compiler settings~\cite{nobre2016graph, kisuki2000combined}. Rejections incur costs, and \SystemName allows users to adjust the significance level to balance benefits against these costs. While \SystemName does not eliminate human intervention, it detects data drift and ageing ML models post-deployment, only requesting user verification when data drift is likely. User feedback is then added to the training dataset to retrain the model.

\cparagraph{Privacy}
Since \SystemName only uses information available to the underlying ML model, it should not raise privacy concerns.

\cparagraph{Integration with non-Python environments}
\SystemName is easy to integrate into non-Python compilers. For example, for C/C++ code, \SystemName provides a pyblind11 API~\cite{pybind11} to take the probabilistic vector of the model prediction as input and returns a boolean value to suggest whether the prediction should be accepted.

\vspace{-1mm}
\section{RELATED WORK}
Supervised ML is a powerful tool for code analysis and optimization~\cite{cummins2017end, POEM, cummins2021programl, zhai2023tlp, Wang2020FUNDED, fu2022linevul, livuldeepecker}. It relies on the assumption that training data will closely resemble future test data~\cite{ribeiro2015mlaas}. However, this assumption can be violated in deployment environments due to evolving hardware and workloads, leading to compromised ML model robustness~\cite{mallick2022matchmaker, singhal2020machine, ackerman2021automatically}. 

Efforts to enhance ML model robustness during design time include data augmentation through code synthesis~\cite{catak2021data}, learning program representations~\cite{cummins2021programl, Wang2020FUNDED}, and tuning ML model architectures~\cite{wang2022automating}. \SystemName complements these design-time methods by improving trained model performance at deployment without altering the model architecture.

Some recent works evaluate DNN prediction accuracy using test data from the operational environment~\cite{guerriero2021operation}, requiring representative data collection from the deployment environment first. \SystemName, however, assesses prediction reliability in real-time. Other methods quantify prediction uncertainties using entropy and mutual information~\cite{gawlikowski2023survey} or train DNN models to estimate prediction confidence for specific applications~\cite{misbehavior}. Unlike \SystemName, which is model-agnostic, these methods depend on specific DNN operators.

\SystemName focuses on detecting incorrect ML predictions caused by changes in input characteristics, combining anomaly detection~\cite{PIMENTEL2014215, RISE, barbero2020transcending} and adversarial attack analysis~\cite{brendel2018decision}. Grounded in Conformal Prediction (CP), guarantees confident predictions. While standard CP libraries like MAPIE~\cite{taquet2022mapie} and PUNCC~\cite{mendil2023puncc} estimate where the ground truth likely lies, \SystemName uses CP differently—assessing the credibility of predictions to reject unreliable ones. Additionally, it offers a framework to automate CP setup and usage.


\SystemName improves upon prior CP methods~\cite{sec2017Transcend, sec2019TESSERACT, xie2020confidence, RISE}, which rely on full calibration datasets and single nonconformity measures. By using adaptive weighting and multiple nonconformity functions, it enhances misprediction detection. Additionally, \SystemName extends CP beyond classification to support regression tasks.

\section{CONCLUSIONS}
We have introduced \SystemName, an open-source library designed to enhance the robustness of learning-based models during deployment for code-related tasks. \SystemName measures the credibility and confidence of predictions to detect likely mispredictions. We evaluated \SystemName on 13 ML models across five code optimization and analysis tasks. \SystemName can effectively detect samples that an ML model can mispredict by successfully identifying an average of 96\% of mispredictions. It also improves deployed model performance by updating the trained model with a few identified test samples. As an example of the practical use of \SystemName, we are working with two companies to integrate PROM into their internal tools to enhance the reliability of ML models for code optimization. 

There is growing interest in applying ML to systems research \cite{wang2018machine, sculley2014machine}, yet prior work has primarily focused on model optimizations during the design phase. \SystemName addresses post-deployment optimization. While this work focuses on code-related tasks, \SystemName can be applied to a wider range of problems with a generic API. We hope that \SystemName will be an enabling technology to support deployment model optimization, thereby enhancing the ML model robustness across various domains.



\vspace{-1mm}
\section*{Data-Availability Statement}
The open-source release of \SystemName is available at \url{https://github.com/HuantWang/PROM}. Additionally, an archival copy of our Artifact can be downloaded from~\cite{artifact} with detailed step-by-step instructions for replicating our results using a Docker image can be found at \url{https://github.com/HuantWang/PROM/blob/ae_cgo/AE.md}.
\vspace{-1mm}
\section*{Acknowledgments}
This work was supported in part by the UK Engineering and Physical Sciences Research Council (EPSRC) under grant agreements EP/X018202/1 and EP/X037304/1.
For any correspondence of this work, please contact Zheng Wang (Email: z.wang5@leeds.ac.uk).

\newpage









\section*{Appendix: Artifacts Evaluation Instructions}

To facilitate the reproduction of our results, we provide a Docker image with a pre-configured environment and detailed README instructions for local evaluation. We assume reviewers are familiar with Python and Linux in a Docker environment. While the Docker image should run on a standard PC, a full-scale evaluation requires a decent multi-core system with sufficient RAM ($\geq$ 64GB) and disk space ($\geq$ 200GB).

\subsection*{Delivery\label{sec:archive}}
We provide a Docker image with step-by-step instructions for running the tool locally, supporting customization and reuse. 
For a step-by-step instruction to replicate our results using a docker image on your machine locally, please refer to
our GitHub repository (\url{https://github.com/HuantWang/PROM/blob/ae_cgo/AE.md}). Our scripts automatically generate numerical results and diagrams corresponding to figures and tables in the paper. 

\subsection*{Main Results}
Our AE enables a reduced-scale evaluation of the main results of our work, specifically Figures 7-11 and Tables 2-3 in Sections 7.1 to 7.5 of the paper. These results show how the tested ML models are impacted by the changes in application workloads and how our framework (PROM) identifies drifting samples to enhance model performance. We compare the detection performance with closely related works. Additionally, we present a small-scale experiment demonstrating the construction of drift detection models to identify mispredictions using our technique and how it can leverage feedback on these samples to improve a deployed model, corresponding to Case 1 of Section 6 of the paper. 

\subsection*{Artifact Check-list (Meta-information)}
\begin{itemize}
\item \textbf{Tool chains:} LLVM v10, Python 3.7 and 3.8 included
\item \textbf{Experiments:} The experiments can be run with the included scripts.
\item \textbf{How much disk space is required (approximately)?:} 200 GB
\item \textbf{How much time is needed to prepare workflow (approximately)? :} Two hours
\item \textbf{Publicly available? :}  Yes, code and data are available at: \url{https://github.com/HuantWang/PROM/}; 

\item \textbf{Code licenses:} CC-BY-4.0 License
\item \textbf{Archived:} 10.5281/zenodo.14077780
\end{itemize}

\bibliographystyle{ACM-Reference-Format}\balance
\bibliography{refs}
\balance

\newpage
\onecolumn
\begin{center}
\textbf{\large Supplemental Material}
\end{center}

\section*{Nonconformity functions\label{appendix:ncm}}
\SystemName currently supports four nonconformity functions that are shown to be useful in prior work~\cite{taquet2022mapie, sadinle2019least, angelopoulos2020uncertainty, romano2020classification, barber2021predictive, kim2020predictive, romano2019conformalized, xu2021conformal}. The nonconformity measures are used to evaluate the credibility and confidence of underlying model predictions, as discussed in Sec. 5.1 of the submitted paper.

\begin{table*}[!ht]
    \caption{Nonconformity functions supported used by \SystemName}
    \label{tab:NCM_formula}
    \scriptsize
    \begin{tabular}{p{1.1cm} p{6cm} p{8cm}}
        \toprule
         \textbf{Functions} & \textbf{Formula} & \textbf{Description} \\
        \midrule
        \rowcolor{Gray} 
        \textit{LAC}~\cite{sadinle2019least} & $1 - \hat{\mu}(X_i)_{Y_i}$ & The conformity score is defined as one minus the score of the true label, where $\hat{\mu}(X_i)_{Y_i}$ is the probability of sample $X_i$ being classified as class $Y_i$. \\
        
         \textit{Top-K}~\cite{angelopoulos2020uncertainty} & $j \quad \text{where} \quad Y_i = \pi_j , \quad \hat{\mu}(X_i)_{\pi_1} > ... > \hat{\mu}(X_i)_{\pi_j} > ... > \hat{\mu}(X_i)_{\pi_n}$ & For a given observation pair $(X_i, Y_i)$, the conformity score $s_i$ is defined as the rank $j$ of the true label $Y_i$ when all the predicted labels $\pi$ are arranged in descending order according to their predicted probabilities $\hat{\mu}(X_i)$. Here, $j$ is the position of the true label $Y_i$ within the ordered set of predicted probabilities, indicating the model's confidence in its prediction.\\
        
        \rowcolor{Gray} \textit{APS}~\cite{romano2020classification} & $\sum^k_{j=1} \hat{\mu}(X_i)_{\pi_j} \quad \text{where} \quad Y_i = \pi_k$ &  The APS method calculates the conformity score $s_i$ for an observation $(X_i, Y_i)$ by summing the predicted probabilities $\hat{\mu}(X_i)_{\pi_j}$ of each label from the highest to the lowest, until the true label $\pi_k$ is reached.\\

         \textit{RAPS}~\cite{angelopoulos2020uncertainty} & $\sum^k_{j=1} \hat{\mu}(X_i)_{\pi_j} + \lambda (k-k_{reg})^+$ & The conformity scores are computed by summing the regularized ranked scores of each label, from higher to lower, until reaching the true label. $\lambda$ is a regularization parameter, and $(k-k_{reg})^+$ is the positive part of the difference between $k$ and the optimal set size $k_{reg}$. \\

        \bottomrule
    \end{tabular}
\end{table*}

\end{document}